\documentclass[a4paper,11pt]{article}
\pdfoutput=1 

\usepackage{jheppub} 
\usepackage{amsmath}
\usepackage[T1]{fontenc} 
\usepackage{subcaption}
\usepackage{xcolor}

\title{
Emergence of Horndeski gravity from asymptotic safety?
}

\author[a]{Astrid Eichhorn,} 
\author[a]{Pedro G.~S.~Fernandes,}
\author[a]{Fabian Willaschek}

\affiliation[a]{Institut f\"ur Theoretische Physik, Universit\"at Heidelberg, Philosophenweg 12 \& 16, 69120 Heidelberg, Germany}

\emailAdd{eichhorn@thphys.uni-heidelberg.de}
\emailAdd{fernandes@thphys.uni-heidelberg.de}
\emailAdd{willaschek@thphys.uni-heidelberg.de}

\abstract{
There are strong motivations to modify gravity in the ultraviolet as well as the infrared. 
Such modifications are usually pursued independently from one another. 
Using the predictive power of asymptotically safe quantum gravity, we can constrain the effective-field-theory coefficients of scalar-tensor theories and thereby connect ultraviolet and infrared modifications of gravity. 
We focus on two non-minimal couplings, which are naturally generated in asymptotic safety and belong to the Horndeski Lagrangian only if the couplings satisfy a specific ratio. A priori, one would expect to find a negative answer to the question in our title. Surprisingly, we find that despite the constraints from asymptotic safety, this ratio can be achieved for a specific value of the cosmological constant. 
We interpret this as a non-trivial hint that asymptotically safe scalar-tensor theories could be free of extra propagating degrees of freedom in the infrared.
\\
We further find that the same result holds in an effective asymptotic safety scenario, where asymptotic safety is not a fundamental theory and quantum scale symmetry, the symmetry underlying asymptotic safety, only holds over an intermediate range of scales.
\\
Finally, we report novel non-trivial indications of approximate radiative stability in the aforementioned Horndeski sector for a range of coupling values, independently of any particular UV completion.
}

\begin{document}
\maketitle
\newpage

\section{Introduction}
General Relativity (GR) is not a complete theory. It needs to be modified in the ultraviolet (UV), in order to account for quantum effects. In response to this need, several candidate quantum gravity theories have been developed. At the opposite end of length scales, there are hints that GR also needs to be modified in the infrared (IR), in order to include additional gravitational degrees of freedom that may model dynamical dark energy.

These two types of modifications are usually pursued independently and largely by ignoring the respective other regime of scales. The reason for doing so is a decoupling argument. It states that, on the one hand, quantum gravity can safely be ignored  at distances larger than the Planck scale, because we cannot dynamically excite the quantum gravitational degrees of freedom that exist at these scales. Conversely, the decoupling argument also states that a low-energy effective field theory (EFT) can be written down without knowledge of the UV physics, and that the unknown UV physics only enters through the free parameters of the EFT.

Yet, a connection of UV and IR aspects of gravity can be very informative, precisely because an EFT has free parameters, which can be constrained or even fixed by a UV theory, enhancing the predictive power of the EFT. 
In turn, the UV theory becomes testable through such predictions of the EFT coefficients. Moreover, what appears as a ``reasonable'' EFT may not be compatible with a given UV completion. This idea is at the heart of the swampland program \cite{Vafa:2005ui,Ooguri:2006in}, in which UV constraints are imposed on otherwise ``reasonable'' EFTs.

In a similar spirit to the swampland program, asymptotically safe quantum gravity \cite{Weinberg:1980gg, Reuter:1996cp}, which is a quantum field theory for gravity, imposes constraints on certain free parameters of the Standard Model of particle physics \cite{Shaposhnikov:2009pv, Harst:2011zx, Eichhorn:2017lry, Eichhorn:2017ylw, Eichhorn:2018whv,Eichhorn:2025sux} and physics beyond the Standard Model, excluding or constraining, e.g., dark-matter candidates \cite{Eichhorn:2017als,Reichert:2019car,Eichhorn:2020kca,Kowalska:2020zve,Hamada:2020vnf,Eichhorn:2021tsx,deBrito:2021akp,Boos:2022pyq,deBrito:2023ydd,Chikkaballi:2025pnw,Assant:2025gto}.
Aside from particle physics, similar efforts have also proven fruitful in cosmology, restricting specific models of dynamical dark energy \cite{Rubio:2017gty,Eichhorn:2022ngh,Wetterich:2024ieb}, cosmic strings \cite{Eichhorn:2023gat}, or inflation \cite{Eichhorn:2020sbo,Silva:2024wit} via asymptotic safety. 
In particular, higher-order interactions, as often invoked for descriptions of dynamical dark energy, are necessarily present in asymptotic safety \cite{Eichhorn:2012va, Eichhorn:2017eht} and can typically be constrained \cite{Eichhorn:2011pc,Christiansen:2017gtg,Eichhorn:2017sok,Eichhorn:2019yzm,Eichhorn:2021qet,deBrito:2021pyi,Laporte:2021kyp,Eichhorn:2022ngh,Eichhorn:2023jyr,deBrito:2023myf,Brenner:2024bps,Eichhorn:2024wba,deBrito:2025nog,Eichhorn:2026euv}.
An up-to-date review of the field of asymptotically safe quantum gravity and its physical implications can be found in \cite{Eichhorn:2026uqj}; further reviews are provided in
\cite{Eichhorn:2018yfc,Pereira:2019dbn,Pawlowski:2020qer,Eichhorn:2022gku,Eichhorn:2022jqj,Eichhorn:2023xee,Wetterich:2022ncl,Knorr:2022dsx,Morris:2022btf,Martini:2022sll,Pawlowski:2023gym,Saueressig:2023irs,Platania:2023srt,Bonanno:2024xne}
and introductions to the field can also be found in two books \cite{Percacci:2017fkn,Reuter:2019byg}.
Important recent progress in the field also addresses the foundational questions of Lorentzian signature and unitarity \cite{Fehre:2021eob,DAngelo:2023wje,Pastor-Gutierrez:2022nki,Pawlowski:2025etp,Saueressig:2025ypi, Kher:2025rve,Chiesa:2026tlz,Knorr:2026vax, Knorr:2026jcg,Assant:2026dca} as well as diffeomorphism invariance \cite{Ihssen:2026ucr}.

In the present paper, we aim to describe a connection between UV and IR properties of gravity, which can (i) constrain the IR properties, (ii) thereby make the UV theory observationally testable (iii) shed light on the structural aspects of the UV theory. 
Within the more general paradigm of UV-IR connections, we focus our efforts largely on asymptotic safety as the UV theory. 
For the IR theory, we consider Horndeski gravity, the most general scalar-tensor theory with second order equations of motion \cite{Horndeski:1974wa}. 
This property guarantees that there are no additional modes beyond the massless spin two mode and a massless or massive scalar mode.
Horndeski gravity is a popular framework to describe dynamical dark energy and its interplay with gravity \cite{Kobayashi:2019hrl}.
\\
The goal of our work is to take a step towards understanding whether Horndeski gravity can emerge as the EFT of asymptotic safety in the IR and whether the UV properties result in constraints on the IR theory. We are motivated by two independent perspectives. First, from a phenomenological perspective, we search for a new theoretical criterion to select among the numerous proposals for modified gravity theories \cite{Clifton:2011jh,Koyama:2015vza,Nojiri:2017ncd,CANTATA:2021asi}. We  advocate that the possibility of emergence from asymptotic safety can act as such a principle. 
Second, our analysis sheds light on a key physical question in asymptotic safety, namely the number and nature of propagating degrees of freedom.
Recent studies of the graviton propagator in asymptotic safety find indications against the existence of additional propagating degrees of freedom, despite the presence of higher-order terms \cite{Pawlowski:2025etp,Knorr:2026jcg,Assant:2026dca}.
We complement such studies from a different perspective, by asking whether a restriction to second-order equations of motion, as built into the Horndeski framework, can effectively be achieved in a theory that exhibits asymptotic safety.

Let us stress that higher-order equations of motion imply the existence of additional degrees of freedom but not necessarily result in problematic instabilities, as is often assumed.
Such an assumption is unwarranted and based on a misinterpretation of Ostrogradsky's theorem. 
The theorem solely states that in point-particle mechanics, higher-order equations of motion, under a non-degeneracy condition, result in a Hamiltonian that is unbounded from above and below. 
The series of works in \cite{Deffayet:2021nnt,Deffayet:2023wdg,Deffayet:2025lnj,Held:2025fii,Deffayet:2026cnu,Deffayet:2026uoe} rigorously proves that, in particular examples, such unboundedness does not result in dynamical instabilities, even at the quantum level. 
What remains of higher-order equations of motion under a non-degeneracy condition are the additional initial conditions required to solve the classical equations of motion. These translate into the number of additional propagating degrees of freedom in the quantum theory \cite{Barnaby:2007ve}.
Whether or not additional degrees of freedom result in problems or pathologies remains to be carefully investigated on a case by case basis, as in \cite{Deffayet:2021nnt,Deffayet:2023wdg,Deffayet:2025lnj,Held:2025fii,Deffayet:2026cnu,Deffayet:2026uoe}. From a purely phenomenological perspective, there are constraints on the number of gravitational degrees of freedom; e.g., there is no evidence that compact binary mergers emit additional degrees of freedom beyond two polarizations of massless spin-2 gravitons \cite{LIGOScientific:2018czr,LIGOScientific:2018dkp}.
\\
With this motivation in mind, we aim to explore whether the constraints on scalar-tensor-theories imposed by demanding asymptotic safety as the UV completion are compatible with the IR constraints defining Horndeski gravity. In this paper, we take a first step in this direction by focusing on dimension six interactions. In the deep IR, we expect that interactions are ordered according to canonical power counting and suppressed by corresponding powers of the Planck mass. Thus, dimension-6-interactions are the critical set of interactions to check.\par\medskip

This paper is structured as follows.
In Sec.~\ref{sec:MainSectionASQGHorndeski}, we address the central question of the paper: whether Horndeski gravity lies in the landscape of asymptotic safety for a massless, $\mathbb Z_2$-symmetric scalar.
To discuss this question, we introduce asymptotic safety and its relation to Horndeski gravity in Sec.~\ref{subsec:IntroASQG}.
Subsequently, in Sec.~\ref{sec:BetaFunctionsIntegration}, we apply the predictive power of asymptotic safety to constrain the scalar couplings.
In Sec.~\ref{sec:Results1}, we present the main result, that a Renormalization Group (RG) trajectory leads to a Horndeski theory in the IR, and explain the mechanism which makes this possible.
In Sec.~\ref{sec:EffectiveASQG}, we relax the requirement of fundamental asymptotic safety and ask how the answer to our question changes when asymptotic safety is an effective theory with a more fundamental theory in the even deeper UV.
Lastly, in Sec.~\ref{sec:RadiativeStability}, we present a general discussion of the radiative stability of the Horndeski coupling relation, independent of any particular UV completion.

\section{Asymptotic safety constraints on non-minimal couplings confront Horndeski gravity}\label{sec:MainSectionASQGHorndeski}

\subsection{Introduction to asymptotic safety}\label{subsec:IntroASQG}

GR is not perturbatively renormalizable as a quantum field theory.
The attempt to do so requires an infinite number of counterterms, each related to a free parameter.
Thus, in the perturbative expansion around vanishing gravitational coupling, one loses all predictive power.
However, perturbative renormalizability is neither sufficient nor necessary for a theory to be UV complete.
The value of a coupling depends on the (energy) scale at which it is probed.
Even when a theory is perturbatively renormalizable, it can feature a Landau pole in the UV, where the coupling diverges at high energies, rendering the theory UV incomplete. One of the most prominent examples of a theory which is perturbatively renormalizable, but not UV complete due to a Landau pole is Quantum Electrodynamics \cite{Gell-Mann:1954yli,Gockeler:1997dn,Gies:2004hy}.
Conversely, a perturbatively nonrenormalizable theory can admit a fixed point in the running of couplings. At such a fixed point, which describes the microscopic UV regime, couplings assume constant, finite values. This allows for a UV completion of the theory and provides a generalized notion of renormalizability because the IR is determined by a typically finite set of relevant interactions. Numerous examples of such interacting fixed points are known; a review of some of them can be found in \cite{Eichhorn:2018yfc}. 
Such a fixed point realizes \emph{quantum scale symmetry}: in the high-energy regime, a change of the energy scale does not further change the coupling values. 
Scale-dependence only sets in at low energies, where the couplings depart from the fixed point and leave the scale-symmetric UV regime.
This is the idea of asymptotically safe quantum gravity (also asymptotic safety for short).
A fixed-point in the running of couplings enables a finite and predictive quantum description of gravity in terms of a quantum field theory.
The advantages of this are twofold:
first, such a theory is highly predictive, leaving only few couplings as free parameters. There is in fact very good evidence for only three such free parameters in asymptotically safe gravity, see, e.g., \cite{Denz:2016qks,Falls:2020qhj,Baldazzi:2023pep} as well as the review \cite{Eichhorn:2026uqj} and references therein.
Second, as a quantum field theory, asymptotically safe quantum gravity is formulated in the same language as many models in modified gravity.
The predictive power of asymptotic safety can thus be directly applied to such models, see, e.g,. \cite{Daum:2010qt,Eichhorn:2013xr,Daum:2013fu,Harst:2014vca,Eichhorn:2022ngh,Gies:2022ikv, Borissova:2025frj,Pastor-Marcos:2026nyb,Heisenberg:2026rdk} for examples.
\par\medskip

By now, asymptotic safety has been investigated using several methods; for an overview, see \cite{Eichhorn:2026uqj}.
The standard method for the analysis of a UV fixed point of gravity is the functional renormalization group (FRG), reviewed in \cite{Dupuis:2020fhh}.\footnote{It is often stated that the FRG is required for the analysis of asymptotic safety as it captures non-perturbative effects, however perturbative studies also support asymptotic safety in gravity, see \cite{Niedermaier:2009zz,Falls:2024noj,Kluth:2024lar}.}
The FRG is based on the Wetterich equation \cite{Wetterich:1992yh,Morris:1993qb,Reuter:1996cp}, a functional differential equation for the scale-dependent effective average action $\Gamma_k$. Solving the Wetterich equation is mathematically equivalent to performing the path integral.
The auxiliary FRG scale $k$ is an infrared (IR) cutoff scale. Quantum fluctuations above this scale are accounted for, whereas quantum fluctuations below the scale are suppressed.
Therefore, the IR limit $k\to0$ of the FRG scale $k$ is the physical limit of the path integral, where all quantum fluctuations are included and $\Gamma_k$ converges to the standard effective action $\Gamma_{k\to0}=\Gamma$, which describes the full quantum theory.
As expected from quantum fluctuations, $\Gamma_k$ contains all terms compatible with the symmetries of the system under investigation, just as the full effective action. The usage of the FRG relies on approximations in the form of truncations of $\Gamma_k$. To obtain physically robust results, truncations can of course not be chosen arbitrarily; instead, a suitable truncation principle needs to be identified, see below.

The scale dependence is encoded in $k$-dependent couplings which parameterize the interaction terms.
The Wetterich equation provides the beta functions 
\begin{equation}
    \beta_{g_i}=k\partial_kg_i,
\end{equation}
which encode the RG running of the couplings.\footnote{We caution that the term ``RG running'' is used for several distinct and inequivalent notions in the literature; this indistinct usage has led to confusion in the past \cite{Donoghue:2019clr},  addressed in \cite{Bonanno:2020bil}; more recent misunderstandings are addressed in \cite{Held:2025vkd}. Here, by RG running, we refer to the dependence of couplings on the scale $k$, which is an unphysical scale and tracks to which extent the path integral has been evaluated. In some situations, this notion of running is in agreement with the physical scale dependence of couplings (e.g., on physical momenta in scattering processes), but in a quantum gravitational context, such different notions of running are in general distinct. See, e.g., \cite{Bonanno:2020bil,Buccio:2024hys,Knorr:2026vax,Chiesa:2026tlz} for further discussions and enlightening examples.}
Given the coupling values at any scale as initial conditions, the beta functions uniquely determine an RG trajectory $g_i(k)$ in theory space.
Every RG trajectory corresponds to a theory with different coupling values.
FRG calculations are often performed in the logarithmic RG scale, which is dimensionless and sometimes called the RG ``time'' 
\begin{equation}
    t=\ln(k/k_0).
\end{equation}
Here, $k_0$ is a reference scale and we note that $\partial_t=k\partial_k$.
\par\medskip

The idea of constructing a theory of quantum gravity through asymptotic safety was first proposed by Weinberg \cite{Weinberg:1980gg}.
The fixed point of asymptotically safe quantum gravity, today called the Reuter fixed point, was first found after the pioneering application of FRG techniques to gravity by Reuter in \cite{Reuter:1996cp,Dou:1997fg,Souma:1999at,Lauscher:2001ya,Reuter:2001ag,Litim:2003vp}.
By today, the Reuter fixed point has been identified in numerous studies, altogether constituting compelling evidence for its existence in Euclidean quantum gravity, see \cite{Eichhorn:2026uqj} and references therein. 
It has been found to have only three free parameters, associated with relevant directions, see \cite{Denz:2016qks,Falls:2020qhj, Knorr:2021slg}. Lorentzian signature is under very active investigation, with promising indications that a change in signature does not affect the existence of the fixed point \cite{Manrique:2011jc, Fehre:2021eob,DAngelo:2023wje,Saueressig:2025ypi,Pawlowski:2025etp,Knorr:2026jcg,Chiesa:2026tlz,Assant:2026dca}. Finally, the effect of matter on the gravitational fixed point has been studied in detail \cite{Dona:2013qba,Meibohm:2015twa,Eichhorn:2015bna,Christiansen:2017cxa,Biemans:2017zca,Alkofer:2018fxj,Wetterich:2019zdo}, as has been the effect of quantum fluctuations of gravity on matter \cite{Shaposhnikov:2009pv,Harst:2011zx,Eichhorn:2017ylw,Eichhorn:2017lry, Eichhorn:2018whv,Alkofer:2020vtb,Pastor-Gutierrez:2022nki,Eichhorn:2025sux}, see \cite{Eichhorn:2022gku, Eichhorn:2026uqj} for reviews and further references.

\subsubsection{The mechanism underlying predictions of asymptotic safety}\label{subsec:PredictionMechanism}

The technical upshot of the mechanism underlying predictions of asymptotically safe quantum gravity is the following:
A fixed point acts as a UV completion for only a subset of RG trajectories, which form a continuous manifold called the critical surface.
To employ the fixed point as a UV completion, we need to restrict the RG flow to this surface.
When the critical surface has a lower dimension than the whole theory space, we can find relations between the couplings that account for this reduction in dimension.
These relations predict some couplings (the irrelevant couplings) in terms of others (the relevant couplings).

To distinguish relevant and irrelevant coupling, let us provide some more details.
In the quantum scale-symmetric regime at the asymptotically safe fixed point, dimensionful couplings $G_i$ change as a function of the energy scale according to their canonical dimension.
It is therefore convenient to work with dimensionless couplings $g_i=k^{-d_{G_i}}G_i$, which are constant in a quantum scale-symmetric regime.
A fixed point  $\vec g_*=\{g_{j*}\}$ of the RG flow is then defined as a zero of the beta function
\begin{equation}
    \beta_{g_i}|_{\vec g_{*}}=0,
\end{equation}
for all $i$.
Expanding the beta function around the fixed point, we find
\begin{equation}\label{eq:BetaFunctionExpandedFP}
    \beta_{g_i}=\sum_j M_{ij}\, (g_j-g_{j*})+O((g_i-g_{i*})^2),
\end{equation}
where $M_{ij}=\partial\beta_i/\partial g_j|_{g_*}$ is the stability matrix.
The zeroth-order term of the expansion vanishes by definition of the fixed point.
We further define the critical exponents $\theta_I$
\begin{equation}
    M_{ij} V_j^I=-\theta_IV_i^I,
\end{equation}
as the eigenvalues of the stability matrix with flipped sign.
Here, $V^I$ are the eigenvectors, which usually constitute a superposition of different couplings $g_i$.
Integrating Eq.~\eqref{eq:BetaFunctionExpandedFP}, we find an approximation for the flow close to the fixed point:
\begin{equation}\label{eq:CouplingExpandedFP}
    g_i(k)=g_{i*}+ 
    \sum_I
    c_I V_i^I\left(\frac{k}{k_0}\right)^{-\theta_I},
\end{equation}
where the $c_I$ are integration constants, and $k_0$ is an arbitrary reference scale.

When the real part of a critical exponent is positive, $\text{Re}(\theta_I)>0$, the associated eigenvector $V^I$ is called a relevant direction.
In that case, the flow out of the UV fixed-point regime into the IR (decreasing $k$), increases the deviation from the fixed-point value in Eq.~\eqref{eq:CouplingExpandedFP}.
The deviation grows until the flow leaves the fixed-point regime.
No prediction can be made for the IR value of a relevant coupling by demanding UV completion through the fixed point; it corresponds to a free parameter.
\\
When the real part of a critical exponent is negative, $\text{Re}(\theta_I)<0$, the associated eigenvector $V^I$ is called an irrelevant direction.
In that case, close to the fixed point, flowing into the IR, the correction to the fixed-point value shrinks, and the flow along that direction converges onto the fixed-point value.
The IR value of an irrelevant direction is thus predicted by the UV completion.
Due to the curvature of the critical hypersurface, the IR prediction of the irrelevant direction is typically not the fixed-point value, but a function of the flow along the relevant directions.
Writing out the irrelevant direction in the basis $\{g_j\}$ leads to a prediction of one of the couplings in terms of the others.
\\
This is the mechanism underlying predictions of asymptotic safety: irrelevant directions reduce the infinitely many dimensions of the theory space to the finite-dimensional critical hypersurface, as only relevant directions are UV completed by the fixed point.\par\medskip

At this stage, one may ask two questions: First, why should there only be finitely many relevant directions? Second, is there a principle that determines which couplings are relevant and which irrelevant?
Both questions are, in fact, connected, because the principle according to which couplings are sorted into relevant and irrelevant also explains why only a few couplings can be relevant, as supported by numerous studies, e.g., \cite{Denz:2016qks,Falls:2020qhj, Knorr:2021slg}.
\\
Critical exponents contain contributions from the canonical dimension of couplings and contributions generated by quantum fluctuations.
A priori, either of the contributions can dominate.
However, an extensive body of work supports the result that in asymptotically safe gravity, as well as in asymptotically safe gravity-matter models, the fixed point is ``near-perturbative'' \cite{Falls:2013bv,Falls:2014tra,Falls:2017lst,Falls:2018ylp,Eichhorn:2018akn,Eichhorn:2018ydy,Eichhorn:2018nda,Eichhorn:2020sbo}. 
This means that the canonical contribution to scaling exponents dominates over the quantum contribution. 
Accordingly, one only needs to inspect the canonical dimension of a coupling to form a well-motivated hypothesis, whether or not it is relevant.
Because only few couplings have positive canonical dimensions, only few couplings are relevant at the Reuter fixed point.
The only exception to this appears to be the Newton coupling, which is canonically irrelevant and must become relevant for a viable fixed point to exist.
When it comes to interactions of matter, there are good indications that even dimension-five interactions, which are the least canonically irrelevant interactions, remain irrelevant in asymptotic safety \cite{deBrito:2021akp,deBrito:2025ges,Assant:2025gto}.
Accordingly, the fixed point is ``as perturbative as it gets''.

Because the canonical dimension remains a good measure for ordering the importance of operators in asymptotic safety (even in the UV), we can construct truncations of the effective action in gravity-matter systems using similar rules as in EFT calculations, which we will use in the next section \ref{subsec:ASST}.
For a given truncation, the near-perturbativity of the fixed point must then be tested explicitly as a consistency check, which we will do in \ref{sec:BetaFunctionsIntegration}.

\subsubsection{Asymptotically safe scalar-tensor theories meet the Horndeski framework}\label{subsec:ASST}

Our main motivation for the scalar field in this work comes from cosmology, where scalar-tensor theories are used in the description of both the early and the late universe. 
We are particularly interested in the late universe, where there are tentative hints from DESI that dark energy is dynamical \cite{DESI:2024mwx,DESI:2025zgx}. This may be described in terms of an additional scalar degree of freedom, non-minimally coupled to gravity \cite{Ye:2024ywg,Wolf:2025jed}.\footnote{A decreasing dark energy component might also arise more directly from quantum gravity, as suggested in causal set quantum gravity \cite{Ahmed:2002mj,Zwane:2017xbg,Das:2023hbw,Das:2023rvg}. However, a more detailed comparison with the observational data from DESI has, so far, not been performed in this setting. It has also been suggested that a dark sector emerges from building blocks of spacetime in quantum gravity \cite{Calcinari:2026xjr}.}
Most model-building efforts of scalar-tensor theories in cosmology start within Horndeski gravity (or some of its extensions, such as beyond Horndeski theory \cite{Gleyzes:2014dya} or DHOST \cite{Langlois:2015cwa,Langlois:2018dxi}), in order to avoid yet further degrees of freedom.\footnote{We caution again that these further degrees of freedom may not lead to catastrophic instabilities. It has been rigorously established that the unboundedness of the Hamiltonian in the presence of higher-order time derivatives does not automatically result in such instabilities \cite{Deffayet:2021nnt,Deffayet:2023wdg,Deffayet:2025lnj,Held:2025fii,Deffayet:2026cnu,Deffayet:2026uoe}. While this does not imply that all higher-order theories are consistent and stable about physically relevant backgrounds, it does imply that the restriction to (beyond) Horndeski gravity  or DHOST may not be necessary to avoid instabilities. It does, however, appear necessary, at least to the best of our knowledge, if no (massive) degrees of freedom should propagate (and the equations of motion are not infinite order).} 
In the standard parametrization, the 
Horndeski action reads \cite{Kobayashi:2011nu}
\begin{equation}
    S_H=\int d^4 x\sqrt{-g}\left[
    \sum_{i=2}^{5}\mathcal L_i
    \right]+S_M,
\end{equation}
where $S_M$ is the matter action, which we neglect in our work. The Lagrange densities are defined as
\begin{align}
\begin{split}
    \mathcal L_2=&G_2(\phi,X),\\
    \mathcal L_3=&G_3(\phi,X)\square\phi,\\
    \mathcal{L}_4 =& G_4(\phi,X)\,R
    +G_{4,X}(\phi,X)\Big[\big(\Box\phi\big)^2-\phi_{;\mu\nu}\phi^{;\mu\nu}\Big],\\
    \mathcal{L}_5 =& G_5(\phi,X)\,G_{\mu\nu}\,\phi^{;\mu\nu}
    -\frac{1}{6}G_{5,X}(\phi,X)\Big[\big(\Box\phi\big)^3
    -3\phi_{;\mu\nu}\phi^{;\mu\nu}\Box\phi
    +2\phi_{;\mu\nu}\phi^{;\nu\rho}\phi_{;\rho}{}^{;\mu}\Big],
\end{split}
\end{align}
where $G_i$ are arbitrary functions of the scalar $\phi$ and we abbreviate the kinetic term as
\begin{equation}
X=-\partial_\mu\phi\partial^\mu\phi/2.
\end{equation}
The semicolons indicate covariant derivatives with respect to the spacetime coordinates, and the commas partial derivatives with respect to the field $\phi$ or the kinetic term $X$, as indicated.
Here, it is our goal to understand whether these phenomenologically motivated frameworks can emerge from the more fundamental perspective of asymptotic safety. At the same time, our study provides information about the number of propagating degrees of freedom in asymptotically safe scalar-tensor theories.
\par\medskip

For a generic scalar-tensor theory to be a Horndeski theory, the couplings in the effective action need to satisfy special relations. These relations ensure a reduction in the order of the equations of motion from higher to second order. They concern interactions with dimension higher than four. Based on the discussion in Sec.~\ref{subsec:PredictionMechanism}, within asymptotic safety, the corresponding couplings are likely to be irrelevant. Accordingly, they are not associated to free parameters and their IR values are expected to be predicted by the UV completion. This makes it highly nontrivial to achieve the desired Horndeski relations. Instead, the general expectation would be that Horndeski gravity does not constitute a good approximation to the effective action of asymptotic safety at low curvature. 
\\
Here, we test this expectation more specifically.
Previous work on the connection of asymptotic safety and Horndeski investigated whether phenomenologically interesting Horndeski coupling values can be achieved from asymptotic safety, finding a negative result \cite{Eichhorn:2022ngh}. While this highlights the predictive power of asymptotic safety for modified gravity, here, we are interested in whether the scalar-tensor theory predicted by asymptotic safety is a Horndeski theory at all.
Our line of reasoning goes as follows: In the effective action $\Gamma_{k\rightarrow 0}$, we consider a derivative expansion. In the presence of massless fluctuations, the effective action generically contains logarithmic terms which may not admit a derivative expansion on a given spacetime background. Within the EFT framework into which Horndeski gravity falls, a derivative expansion is typically assumed to exist. We adopt this assumption here.
Starting from a near-perturbative asymptotically safe fixed point, one expects that higher-order interactions in this expansion feature additional suppression by a positive power
of the ratio of the curvature/energy scale divided by the Planck-scale. 
Accordingly, we focus on the lowest-order interactions in the effective action which lie beyond Horndeski gravity.\footnote{In a purely gravitational setting, the first such couplings are those of quadratic gravity, which are associated with one relevant direction for the two couplings in asymptotic safety.
We leave the treatment of these terms for future work to focus on scalar-tensor theories in this work. Let us also highlight that in studies of the graviton propagator beyond a derivative expansion, no poles appear as they would be associated to quadratic curvature couplings \cite{Pawlowski:2025etp,Knorr:2026jcg,Assant:2026dca}. In fact, \cite{Knorr:2026jcg} even finds support for a non-trivial mechanism \cite{Platania:2020knd} that removes additional poles and thus propagating degrees of freedom. This mechanism relies on infinitely many higher-order terms, which, when expanded to finite order, give rise to expansion-induced poles.}
Whether the corresponding couplings satisfy a particular relation is a crucial test of whether the effective action of asymptotically safe scalar-tensor theories is well-approximated by the Horndeski action in the IR. We expect the answer to be negative because previous studies have found both couplings, that we introduce below, to be irrelevant \cite{Eichhorn:2017sok, Laporte:2021kyp}.

To study these questions, we restrict ourselves to scalar-tensor theories without further matter fields. In a more realistic setting, Standard Model degrees of freedom are also present. In phenomenological model building, couplings between a dark energy sector and the Standard Model matter fields are usually set to zero. In asymptotic safety, gravitational fluctuations induce such couplings, which means they cannot generally be set to zero \cite{Eichhorn:2012va, Eichhorn:2017eht}. 
However, for our scalar field in the dark sector, such induced interactions with Standard Model fields are dimension-eight or higher, and by previous arguments likely negligible for our purposes.
\par\medskip

Within scalar-tensor theories, we focus on those interactions that are unavoidably generated by quantum gravity fluctuations.
As soon as we have a scalar kinetic term, quantum fluctuations of asymptotic safety generate all possible terms invariant under the two global symmetries of the kinetic term, namely $\mathbb Z_2$ symmetry $(\phi\to-\phi)$ and shift symmetry $(\phi\to\phi+c)$ \cite{Eichhorn:2012va, Eichhorn:2017eht, Eichhorn:2017sok,Laporte:2021kyp,Eichhorn:2020sbo,deBrito:2021pyi, deBrito:2023myf}. 
In terms of operator order, the first terms generated by this mechanism are the two dimension-six non-minimal derivative operators $R^{\mu\nu}\partial_\mu\phi\partial_\nu\phi$ and $R\, g^{\mu\nu}\partial_{\mu}\phi\partial_{\nu}\phi$.
Accordingly, our truncation reads
\begin{equation}
\label{eq:OurTruncation}
\Gamma_k=\Gamma^{\text{EH}}_k+\Gamma^{\text{min}}_k+\Gamma^{\text{nonmin}}_k,
\end{equation}
with the gravitational Einstein-Hilbert term
\begin{equation}
    \Gamma^{\text{EH}}_k=\int d^4x\sqrt{\det g_{\mu\nu}}\left(
    \frac{2\Lambda(k)-R}{16\pi\, G(k)}
    \right),
\end{equation}
the minimally coupled scalar kinetic term
\begin{equation}
    \Gamma^{\text{min}}_k=\int d^4x\sqrt{\det g_{\mu\nu}}\left(
   \frac{1}{2}
   g^{\mu\nu}\partial_{\mu}\phi\partial_{\nu}\phi
    \right),
\end{equation}
and the non-minimal dimension-six operators which close our truncation
\begin{equation}
    \Gamma^{\text{nonmin}}_k=\int d^4x\sqrt{\det g_{\mu\nu}}\big(
    C_{T}(k) R^{\mu\nu}\partial_\mu\phi\partial_\nu\phi
    +C_{S}(k) R\, g^{\mu\nu}\partial_{\mu}\phi\partial_{\nu}\phi
    \big).
\end{equation}
We write out the spacetime volume factor explicitly to avoid confusion with the gravitational coupling $g$ and further employ Euclidean signature, as is standard in most FRG calculations.
Our truncation contains four couplings: the cosmological constant $\Lambda$, the gravitational coupling $G$ and the two non-minimal scalar couplings $C_S$ and $C_T$. We extract the $k$ dependence of their dimensionless counterparts from the Wetterich equation.

In App.~\ref{app:ExtendedTruncation} we will extend the truncation in Eq.~\eqref{eq:OurTruncation} to test the robustness of our results. To do so, we include the running of the wave-function renormalization of the scalar, as well as the dimension-eight interaction that constitutes the square of the kinetic term.

In the FRG, the effect of the non-minimal derivative coupling to the Ricci scalar was first investigated in \cite{Eichhorn:2017sok}, and the beta functions of the whole truncation in Eq.~\eqref{eq:OurTruncation} as well as the extended truncation in App.~\ref{app:ExtendedTruncation} have been calculated in \cite{Laporte:2021kyp}, following studies of the dimension-eight selfinteraction in \cite{Eichhorn:2012va, deBrito:2021pyi}.
Other interaction terms for scalar fields, both minimal and non-minimal ones, have been studied in \cite{Narain:2009fy,Eichhorn:2012va,Labus:2015ska,Percacci:2015wwa,Oda:2015sma,Dona:2015tnf,Eichhorn:2017sok,Pawlowski:2018ixd,Wetterich:2019rsn,Eichhorn:2020sbo,deBrito:2021pyi,Ohta:2021bkc,Knorr:2022ilz}.

The physical effective action is achieved in the limit $k \rightarrow 0$, where consequences, e.g., for cosmology could be extracted. Thus, our main aim in this work is calculate $\Gamma_{k \rightarrow 0}$ in the above truncation, starting from an asymptotically safe fixed point. We caution that the $k$-dependence is a priori unphysical and simply constitutes an auxiliary scale dependence that reorganizes the path integral into a collection of differential equations, the beta functions.
\par\medskip

In terms of the Horndeski action in Eq.~\eqref{eq:ClassicalHorndeskiAction}, our truncation amounts to 
\begin{equation}
    G_4=\frac{1}{16\pi G}+CX,\quad G_2=X-\frac{\Lambda}{\kappa^2}.
\end{equation}
Using commutator identities of the covariant derivative, we can rewrite the resulting Horndeski theory in terms of the Einstein tensor
\begin{equation}
G_{\mu\nu} = R_{\mu\nu} - \frac{1}{2}g_{\mu\nu}R,
\end{equation}
to obtain\footnote{The same action can be constructed using $G_5=-C\phi$.}
\begin{equation}\label{eq:ClassicalHorndeskiAction}
    S_H=\int_x\sqrt{-g}\left(
    \frac{R-2\Lambda}{16\pi G}+CG_{\mu\nu}\partial^\mu\phi\partial^\nu\phi-\frac{1}{2}(\partial\phi)^2
    \right).
\end{equation}
As this is a classical IR theory, here, couplings are constant and do not carry $k$-dependence.
Compared to the FRG truncation in Eq.~\eqref{eq:OurTruncation}, they correspond to a physical limit $k\to0$. 
We find that the expressions match if
\begin{equation}
C_T(k)\xrightarrow{k\to0}C \quad \mbox{and}\quad C_S(k)\xrightarrow{k\to0}-C/2,
\end{equation}
for an arbitrary constant C.
In other words: the truncated effective action implied by asymptotic safety is a Horndeski theory exactly when, in the physical IR limit, the two non-minimal operators $C_T R^{\mu\nu}$ and $C_S Rg^{\mu\nu}$ combine to form the Einstein tensor.\footnote{If this is not the case and the theory is not Horndeski, it is also not part of DHOST.}
This amounts to the condition
\begin{equation}\label{eq:HorndeskiCondition}
    \boxed{C_{S}/C_{T}\xrightarrow{k\to0}-1/2}.
\end{equation}
\par\medskip

Horndeski gravity is usually discussed as a classical modified theory of gravity beyond GR.
We repeat that, in contrast to what is often done in the literature in cosmological studies, in our case the effective action \eqref{eq:OurTruncation} is not the result of model-building.
Instead, the values of the couplings $C_T$ and $C_S$ in the effective action \eqref{eq:OurTruncation} are implications of asymptotic safety. Therefore, we cannot simply adjust $C_T$ and $C_S$ as needed, in order for the action to be part of the Horndeski framework.
\\
From the perspective of a high energy quantum theory, such as asymptotic safety, Horndeski gravity is one particular set of specific low-energy (IR) effective field theories.
Formulated in that language, we are interested in the question of whether Horndeski gravity lies in the ``landscape'' or the ``swampland'' of asymptotic safety.
The landscape of a UV theory is the set of low-energy EFTs that are compatible with the more fundamental UV theory as their UV completion.
Conversely, the swampland consists of those EFTs which are not compatible with the more fundamental theory as UV completion.
The idea of the swampland program comes from string theory \cite{Vafa:2005ui,Palti:2019pca} but has recently been expanded to asymptotic safety \cite{Basile:2021krr,Basile:2025zjc,Eichhorn:2024rkc,Knorr:2024yiu}.
If an EFT lies in the ``relative swampland'', i.e., in the swampland of one UV theory but in the landscape of another, an observational constraint on the EFT amounts to an indirect test of quantum gravity \cite{Eichhorn:2024rkc}.

In the Horndeski literature, the non-minimal derivative coupling to the Einstein tensor in \eqref{eq:ClassicalHorndeskiAction} is known as the "John" term of the "Fab Four" \cite{Charmousis:2011ea,Charmousis:2011bf}.
The Fab Four subset of Horndeski theories can lead to large-$\Lambda$ self-tuning of the cosmological constant on FLRW backgrounds.
Such a scenario addresses the fine-tuning problem of the cosmological constant.
This has further been upgraded to the "Fab Five" which also allow for the inclusion of a kinetic term and works on de Sitter backgrounds \cite{Appleby:2012rx,Appleby:2015ysa,Starobinsky:2016kua}.

As a model of late time cosmology, the non-minimal derivative coupling to the Einstein tensor is strongly constrained by measurements of the speed of gravitational waves \cite{LIGOScientific:2017vwq,Ezquiaga:2017ekz}, although a frequency dependence of the speed of gravitational waves might limit the constraining power \cite{deRham:2018red}.
Here, we consider this coupling as an EFT correction implied by asymptotic safety.
This means that the couplings $C_S$ and $C_T$ feature Planck-suppression according to their dimensions.
They thereby easily fulfill the observational constraints.
\par\medskip

The comparison between the results from asymptotic safety and the Horndeski framework is based on two important approximations.
First, the Horndeski action in Eq.~\eqref{eq:ClassicalHorndeskiAction} is formulated in Lorentzian signature, while the FRG truncation in Eq.~\eqref{eq:OurTruncation} is in Euclidean signature.
Comparing the two is based on the assumption that the two actions are connected by a Wick rotation.
By now, there are first FRG studies in Lorentzian signature as well as analytical continuation \cite{Fehre:2021eob,Kher:2025rve,Pawlowski:2025etp,Assant:2026dca,Knorr:2026jcg}, which suggest that an asymptotically safe fixed point persists in Lorentzian signature. 
Here, we make the assumption that we can simply analytically continue at the level of the action and use the Euclidean RG flows to calculate the corresponding low-energy values of the couplings in the Lorentzian action.
A short calculation shows that a simple Wick rotation amounts to a sign change in both $C_S$ and $C_T$.
This leaves the sign of their ratio invariant.
\\
Second, the truncation in Eq.~\eqref{eq:OurTruncation} only includes the \textit{dominant} non-minimal coupling of a \textit{shift- and $\mathbb Z_2$-symmetric} scalar.
Even under these symmetries, including higher-order operators in the truncation, e.g., $R^2(\partial\phi)^2$, necessarily leads to higher-order equations of motion, and one generally cannot fulfill the Horndeski property.
However, these terms carry additional suppression by ratios with the Planck scale, and the truncation in Eq.~\eqref{eq:OurTruncation} should constitute a good approximation.
More precisely, under the condition Eq.~\eqref{eq:HorndeskiCondition}, the low-energy EFT of a massless asymptotically safe scalar-tensor theory is \textit{well-approximated} by a Horndeski theory, as long as we can additionally neglect the quadratic-gravity terms, which we assume here.
Relaxing the shift-symmetry to include a scalar mass and non-derivative non-minimal couplings leads to a different system which could qualitatively change the analysis, especially when the mass becomes large.
This is left for future work.

\subsection{Beta functions, fixed point and integration towards the infrared}\label{sec:BetaFunctionsIntegration}

In the following, we will use the FRG beta functions from Ref.~\cite[Eqs.~(4.25)-(4.27)]{Laporte:2021kyp}.\footnote{The truncation in Eq.~\eqref{eq:OurTruncation} excludes the square of the kinetic term and the wave function renormalization of the scalar.
In the notation of the paper \cite{Laporte:2021kyp}, this corresponds to $c=0$ and $\eta_S=0$. They call this system "$\tilde cd$" (without a "*").
The coupling called $C_T$ in this text corresponds to $\tilde C$ in \cite{Laporte:2021kyp}, and $C_S$ to $D$. We include the additional couplings in the analysis in App.~\ref{app:ExtendedTruncation}.
}
Using the beta functions, we need to identify a suitable UV fixed point for the dimensionless couplings, denoted by lower-case letters
\begin{equation}
    G=g/k^2,\quad \Lambda=\lambda k^2,\quad C_T=c_T/k^2,\quad C_S=c_S/k^2.
\end{equation}
After inserting the threshold functions using the Litim regulator \cite{Laporte:2021kyp,Reuter:2001ag}, we solve for the zeros of the beta functions. 
The zeros of the beta functions are fixed-point candidates, which need to be restricted to physically viable fixed points according to their critical exponents.
When the difference between the classical dimension of the couplings and their critical exponents is too large, the fixed point is not ``near-perturbative''.
Such a fixed-point candidate breaks the self-consistency of the truncation and should accordingly be discarded.
In this truncation, we find exactly one fixed point with suitable critical exponents.
It reads\footnote{We provide results rounded to two decimals. While the numerical accuracy is much higher, this reflects the fact that it is difficult to exactly pinpoint the systematic error of the truncation, which we do not analyze in detail in this work.
We further note small differences in the values for the fixed point and critical exponents compared to \cite{Laporte:2021kyp}.
}
\begin{equation}\label{eq:OurFP}
   (g,\lambda,c_T,c_S)=(0.66,\,0.22,\,0.36,-1.28)=:\text{FP},
\end{equation}
and has critical exponents
\begin{align}\label{eq:CritExponents}
    \theta_{1,2}&=1.48 \pm 3.56 i,\\
    \theta_{3,4}&=-2.56 \pm 0.76 i,
\end{align}
with associated eigenvectors
\begin{align}\label{eq:Eigenvectors}
    V_{1,2}&=(-0.32\pm0.24i,\, -0.05\mp 0.11i,\, -0.22\pm0.18i,\ 0.87),
    \\ \label{eq:Eigenvectors34}
    V_{3,4}&=(\mathcal{O}(10^{-3}),\,\mathcal{O}(10^{-3}),\, 0.86,\,-0.02\mp0.51i ).
\end{align}
In the following, we will discuss the fixed point in Eq.~\eqref{eq:OurFP} as the UV completion of the gravity-scalar system in Eq.~\eqref{eq:OurTruncation}.
The two critical exponents $\theta_{1,2}$ have a positive real part and thus $V_{1,2}$ correspond to relevant directions. In contrast, $\theta_{3,4}$ have a negative real part, such that $V_{3,4}$ correspond to irrelevant directions.
We note that -- as is standard at interacting fixed points -- the eigenbasis of the stability matrix, i.e., the set of eigenvectors, is not aligned with the original basis, in which each coupling is associated to a basis element.
Converting from the eigenbasis back to the original coupling basis, the UV completion through the fixed point in Eq.~\eqref{eq:OurFP} allows us to predict the values of $(c_S,c_T)$ as functions of the values of $(g,\lambda)$.
We exploit this in Sec.~\ref{subsec:LeveragingPredictivePowerASQG}.
\par\medskip

As discussed in Sec.~\ref{subsec:PredictionMechanism}, the critical surface, spanned by the relevant directions, describes the part of the theory space that is UV-completed by the fixed point in Eq.~\eqref{eq:OurFP}. We will now map out the critical surface.\footnote{For other examples and discussions of how to achieve such mappings, see, e.g., \cite{Saueressig:2024ojx,Knorr:2024yiu,Eichhorn:2025xbb}.}
The physically meaningful direction of the flow is from UV to IR: the microphysics determines the macrophysics.
We thus initiate the flow starting at a small distance $\epsilon$ away from the fixed point, along (a superposition of) relevant directions.
An initial point in the critical surface, together with the beta functions, uniquely determines the flow.
The initial distance $\epsilon$ must be chosen small enough for the initial point to be in the linearized regime around the fixed point, where the critical hypersurface is well-approximated by a plane with vanishing (extrinsic) curvature.
A choice of small $\epsilon$ is related to the choice of units and the relative scale $k_0$.
The two-dimensional critical surface is thus spanned by a single angle $\alpha$ and we initiate the flow at
\begin{equation}\label{eq:InitalPointFlow}
    P_{}(\alpha)=\text{FP}+\epsilon\big(
    \cos(\alpha)\widetilde V_1+\sin(\alpha)\widetilde V_2
    \big).
\end{equation}
Here, $\widetilde V_{1}=V_1+V_2$ and $\widetilde V_{2}=i(V_1-V_2)$ are the real projections of the relevant eigenvectors $V_{1,2}$, as the coupling values must remain real.
$\alpha$ is the angle of exit from the fixed point. 
Given an initial point described by Eq.~\eqref{eq:InitalPointFlow}, parametrized by an angle $\alpha$, the flow interpolates between the UV fixed point (large $k$) and the resulting IR behavior (small $k$), which we will analyze in the following.
\par\medskip

\begin{figure}
    \centering
    \includegraphics[width=0.6\linewidth]{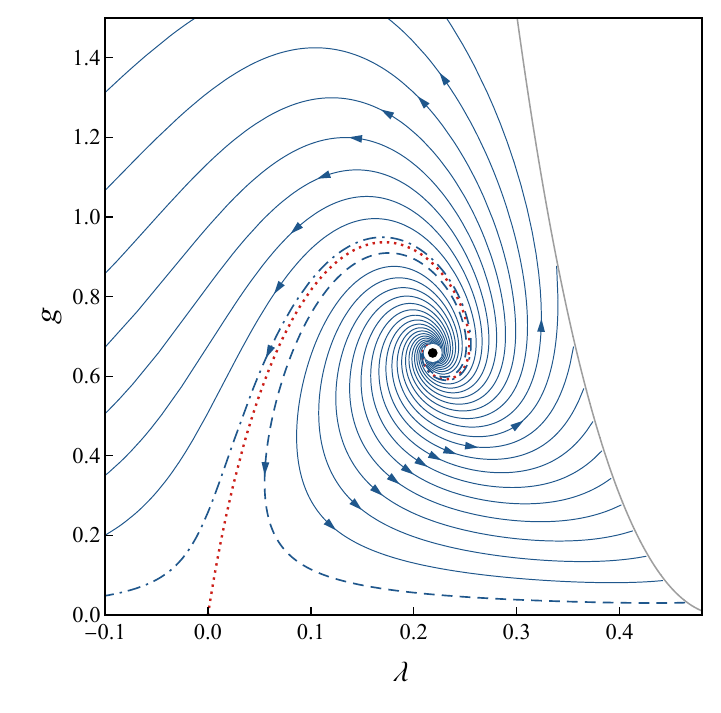}
    \caption{We show the RG flow of the $(g,\lambda,c_T,c_S)$-system from the UV fixed point to the IR, projected onto the $(g,\lambda)$-plane.
    The result is the well-known rotating flow around the Reuter fixed point \cite{Reuter:2001ag}.
    Here, we show the flow of 20 equally spaced angles of exit from the fixed point, as defined in Eq.~\eqref{eq:InitalPointFlow}.
     The dotted red line is the separatrix, which is found by fine-tuning the angle of exit.
    We highlight one dS (dashed) and one AdS (dash-dotted) trajectory close to the separatrix as examples.
    The black line corresponds to the position of the off-shell pole where the integration breaks down.
    }
    \label{fig:glambdaStreamPlot_0-2pi}
\end{figure}

We first discuss the flow in the $(g,\lambda)$-plane.
The flow resulting from 20 evenly spaced angles is displayed in Fig.~\ref{fig:glambdaStreamPlot_0-2pi}.
Qualitatively, three types of flows can be distinguished by their IR behavior, as already discussed in \cite{Reuter:2001ag}.
The first type of flow goes towards positive cosmological constant (dS).
The corresponding trajectories
stop at a finite, positive, g-dependent value of $\lambda$.
This is the so-called lambda-1/2 singularity, which is not a physical phenomenon, 
but a result of the off-shell nature of our calculation, see, e.g., \cite{Percacci:2015wwa,Knorr:2021slg}.
The IR behavior of trajectories that run into the lambda-1/2 singularity cannot be evaluated from this calculation.
The second type of trajectories goes to negative cosmological constant (AdS).\footnote{Some of the trajectories that run into the lambda-1/2 singularity might also be AdS paths that stop before reaching negative $\lambda$.}
These trajectories can be evaluated at $k\rightarrow 0$, where they behave as $\lambda \rightarrow - \infty$, as required for the dimensionful cosmological constant $\Lambda=\lambda k^2$ to converge to a finite value.
The third type of trajectory, between the other two cases, is the separatrix, which flows towards the Gaussian fixed point $(g=0,\lambda=0)$ and leads to a vanishing cosmological constant.

\subsubsection{Leveraging the predictive power of asymptotic safety}\label{subsec:LeveragingPredictivePowerASQG}

To reproduce the Newtonian limit, the dimensionful gravitational constant $G$ has to converge to Newton's constant $G_N$ in the IR
\begin{equation}\label{eq:GNinIR}
    G=gk^{-2}=\frac{g}{k_0^2}e^{-2t}\xrightarrow{t\to-\infty}1=G_N=m_P^{-2},
\end{equation}
where $m_P=\sqrt{8\pi}M_P$ is the non-reduced Planck mass, and by setting $G_N=1$ we have chosen natural units.
After integrating the system in $t=\ln k/k_0$, the relation \eqref{eq:GNinIR} can be used to find the relative scale $k_0$ corresponding to our choice of units. Here, this is not required as we only analyze dimensionless coupling combinations.
The limit $t=\ln(k/k_0)\to-\infty$ is equivalent to the limit $k\to0$.
\\
To set units for the other couplings, we need to evaluate them relative to $g$.
This creates dimensionless ratios which, in the IR, converge to the value of the dimensionful coupling in natural units.
For example, to find the (physical) IR cosmological constant in natural units, we have to evaluate $g\lambda$ in the IR:
\begin{equation}
    \lambda g=(\lambda k^2)(g/k^2)=\Lambda G\xrightarrow{t\to-\infty}
    \Lambda_{\text{IR}} G_N=\Lambda_{\text{IR}}\,[l_P^{-2}].
\end{equation}

\begin{figure}
    \centering
    \includegraphics[width=\linewidth]{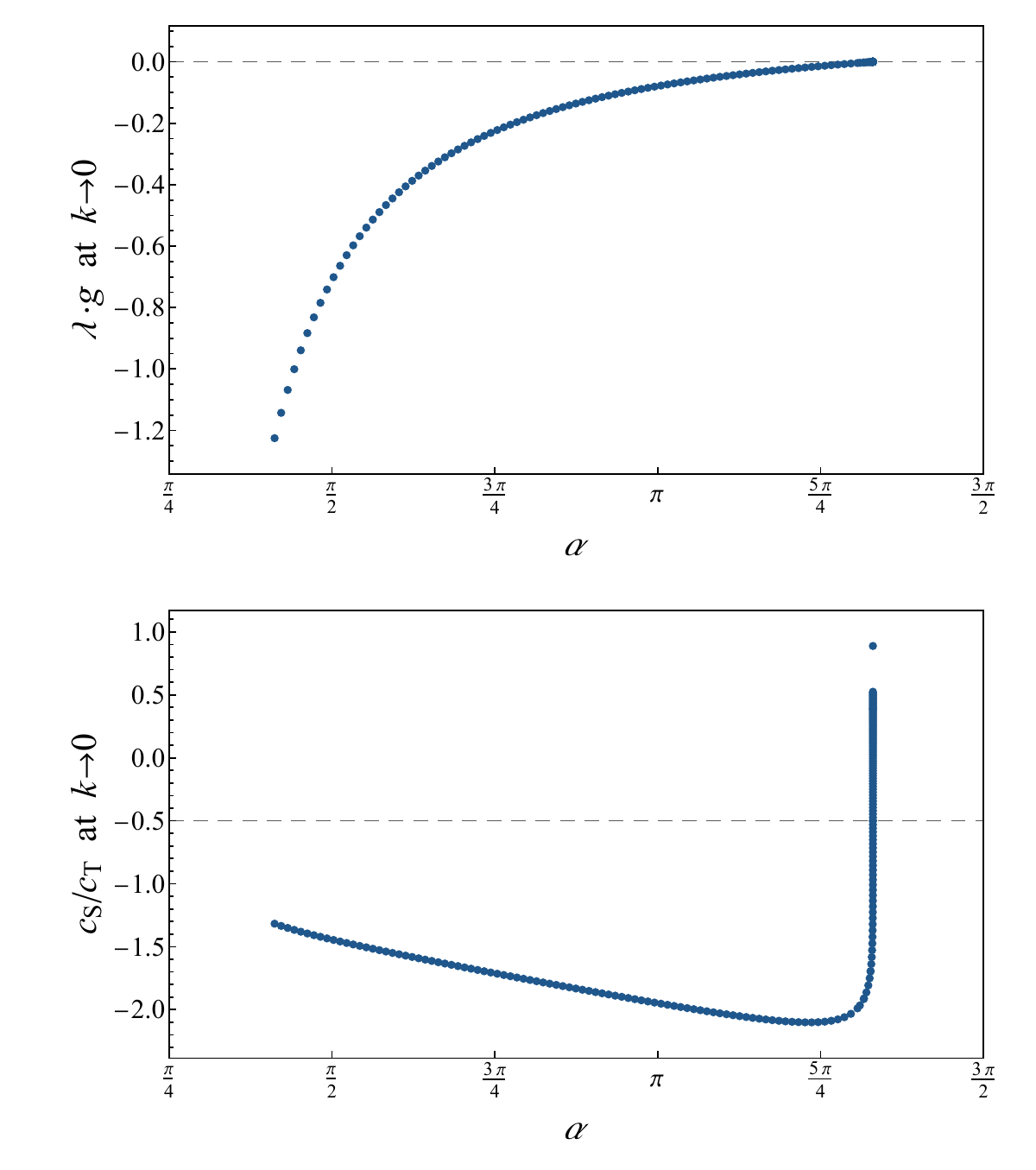}
    \caption{
    We show the physical ($k\to0$) values of the dimensionless coupling combinations $g\lambda$ (upper panel) and $c_S/c_T$ (lower panel) as a function of the angle of exit $\alpha$ from the UV fixed point. 
    We only include the endpoints of trajectories that converge for $k \rightarrow 0$, which is why $\lambda$ is negative for all cases.
    The IR values of $c_S/c_T$ exhibit strong growth close to the separatrix, where $g\lambda\to0$.
    To resolve the steep behavior of the scalar coupling ratio close to the separatrix, this includes fine-tuned angles in addition to a scan in even steps.
    The trajectories cross the Horndeski value of $c_S/c_T=-1/2$.
    The ratio $c_S/c_T$ is bounded by 2 exactly on the separatrix, cf.~Sec.~\ref{subsec:FlowRegimes}. 
    }
    \label{fig:IRValues02pi}
\end{figure}

On this basis, we can explore which values of the cosmological constant are possible in the IR limit.
Our results are displayed in the upper panel of Fig.~\ref{fig:IRValues02pi}.
Each of the data points corresponds to a universe with a different value of the cosmological constant.
We observe that the closer a trajectory is to the separatrix, the smaller its absolute value of the cosmological constant in the IR.
This behavior will be explained in Sec.~\ref{subsec:FlowRegimes} and is the incarnation of the cosmological-constant fine-tuning problem within the functional RG formalism.\footnote{This problem may be circumvented within unimodular gravity \cite{Weinberg:1988cp}, in which evidence for asymptotic safety exists \cite{Eichhorn:2013xr,Eichhorn:2015bna,Benedetti:2015zsw,DeBrito:2019gdd,deBrito:2020xhy,deBrito:2022vbr}. Repeating a similar analysis of Horndeski coupling ratios in a unimodular setup is therefore an interesting extension of our work, left for the future.}
\\
In summary, we have now reduced the two free parameters, corresponding to $g$ and $\lambda$, to one physical free parameter, whereas the other one is used to define units. We parameterize this free parameter as the angle of exit from the fixed point. 
On this basis, we can determine the scalar couplings $c_S$ and $c_T$.
To obtain the corresponding dimensionful couplings in Planck units, we again evaluate the couplings relative to $g$ in the IR limit $k \rightarrow 0$:
\begin{equation}
    \frac{c_{S/T}}g=\frac{C_{S/T}}G\xrightarrow{t\to-\infty}C_{S/T,\,\text{IR}}\,[m_P^{-2}].
\end{equation}
In the ratio $C_S/C_T=\frac{C_S}{G}/\frac{C_T}{G} =c_S/c_T$, the gravitational coupling cancels out, as the ratio is dimensionless.
To search for the Horndeski ratio $C_S/C_T=-1/2$, we can therefore directly work with $c_S/c_T$; see the lower panel of Fig.~\ref{fig:IRValues02pi}.

\subsection{Results: Horndeski coupling ratio for universes with tiny $|\Lambda|$}\label{sec:Results1}

\begin{figure}[!t]
        \begin{center}
    
    \begin{subfigure}{\linewidth}
        \centering
        \includegraphics[width=0.55\linewidth]
        {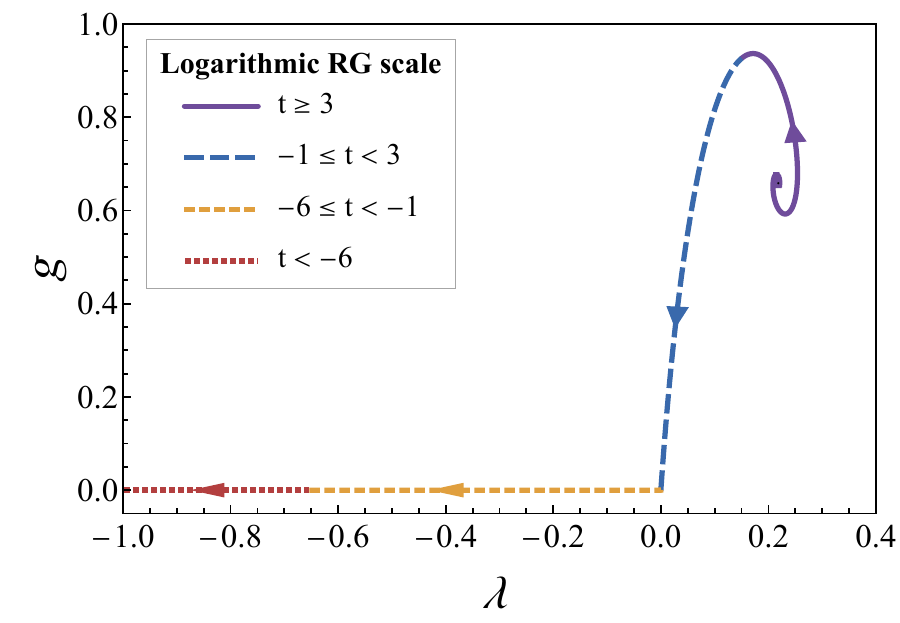}
        \caption{Horndeski trajectory in the $(\lambda,g)$-plane. The distinct colors indicate distinct regimes of the flow, discussed in Sec.~\ref{subsec:FlowRegimes}.}
        \label{fig:StreamPlotHorndeski}
    \end{subfigure}

    \vspace{0.7em}

    \begin{subfigure}{\linewidth}
\centering
        \includegraphics[width=\linewidth]{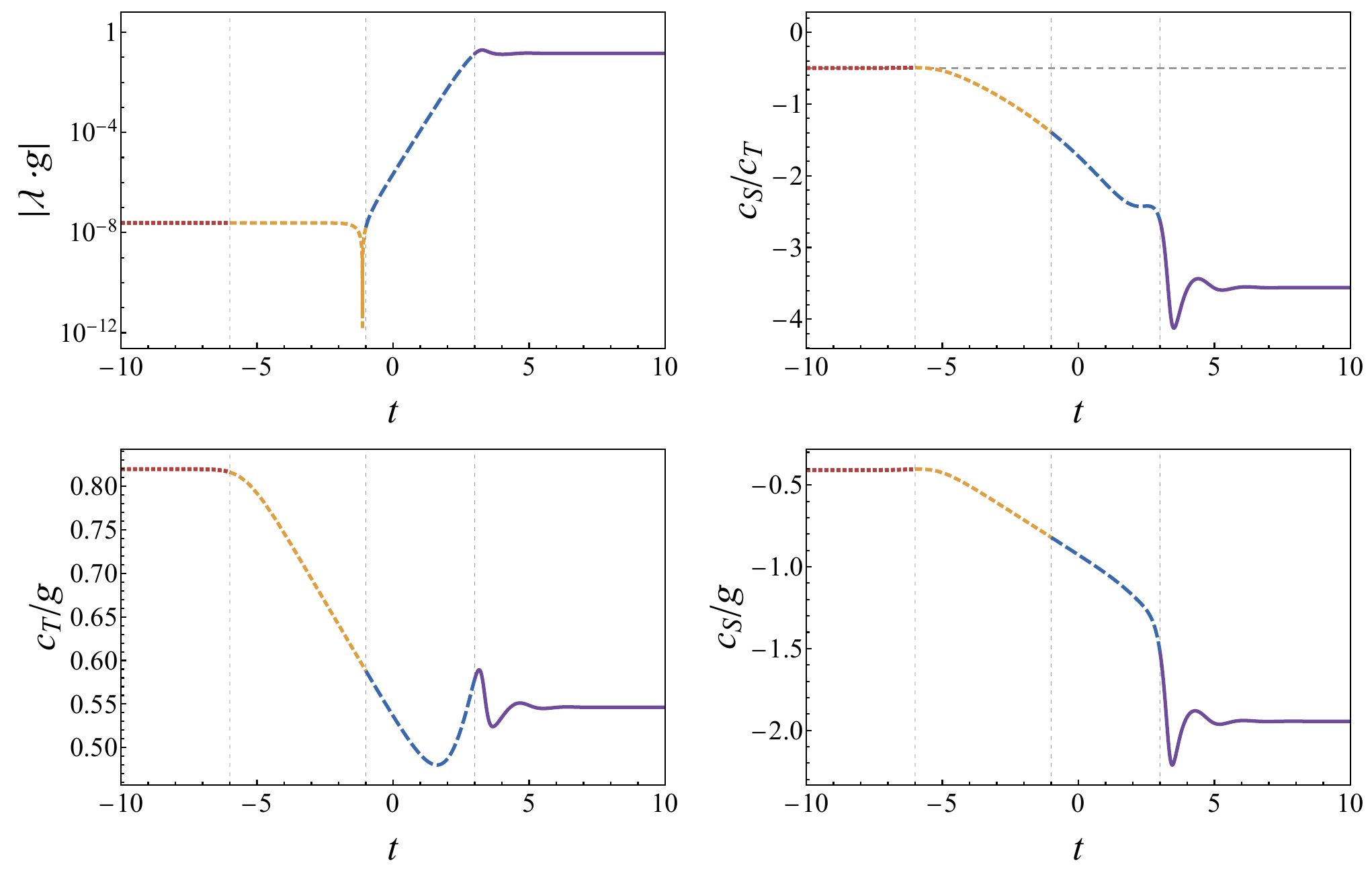}
        \caption{Running dimensionless coupling combinations $|\lambda g|$, $c_S/c_T$, $c_S/g$, and $c_T/g$ of the Horndeski trajectory as functions of the logarithmic RG scale (``RG time'') $t=\ln(k/k_0)$.
        The specific value of the reference scale $k_0$ is not of importance for this discussion.
        The magnitude $\lvert\lambda g\rvert$ is displayed on a logarithmic scale to highlight the Gaussian scaling.
        The ratios $c_S/g$ and $c_T/g$ are displayed on linear scales, as they scale linearly in $t$ (logarithmically in $k$) close to the Gaussian fixed point; see Sec.~\ref{subsec:FlowRegimes}.
        In the top-right panel, we confirm that this trajectory converges to the Horndeski ratio
        $c_S/c_T\to-1/2$ (horizontal dashed line) in the IR.
        }
        \label{fig:RGflowCouplingsHorndeski}
    \end{subfigure}
\end{center}
    \caption{We show the RG trajectory that realizes the Horndeski ratio.}
    \label{fig:HorndeskiRGFlow}
\end{figure}

In Fig.~\ref{fig:IRValues02pi}, we identify a single converged RG trajectory that leads to the Horndeski ratio $c_S/c_T=-1/2$. We emphasize that this is a highly non-trivial result because, in general, it is not expected that asymptotically safe scalar-tensor theories can satisfy the Horndeski constraints, given that there are no free parameters associated with the non-minimal couplings.
In terms of the $(g,\lambda)$ plane, the Horndeski trajectory is very close to the separatrix (the red-dotted line in Figure~\ref{fig:glambdaStreamPlot_0-2pi}), before it branches out to the left close to the Gaussian fixed-point, see Figure~\ref{fig:StreamPlotHorndeski}.
We will delve into more details on the mechanism behind this result in Sec.~\ref{subsec:FlowRegimes}. Here, we only point out that the mechanism is possible, because one free parameter, associated to the cosmological constant, is ``sacrificed'' to tune $c_S/c_T=-1/2$, rather than the value of $\Lambda$. 
Consequently, we obtain a Horndeski theory in the IR only for a specific value of the cosmological constant. 

In the physical limit $k \rightarrow 0$ of the Horndeski trajectory we find the following values for the dimensionful couplings
\begin{equation}
\Lambda_{\text{IR}}\approx-2.42\times10^{-8}\,[l_P^{-2}],\quad C_{T,\mathrm{IR}}\approx0.82\,[m_P^{-2}],\quad  C_{S,\mathrm{IR}}\approx-0.41\,[m_P^{-2}].
\end{equation}
The resulting cosmological constant is a relatively small number in natural units. This may make it more plausible that a self-tuning mechanism, e.g., along the lines of mechanisms proposed within the ``Fab Five'' can be invoked. Such a mechanism would be needed, because our value for $\Lambda$ not only has the wrong sign, but is still many orders of magnitude larger than the value $10^{-122}$ expected in $\Lambda$-CDM \cite{Planck:2018vyg}.
We emphasize that $C_T$ and $C_S$ are now fixed individually; the Horndeski ratio can \emph{not} be attained for arbitrary values of $C_S$ and $C_T=-2 C_S$. Accordingly, asymptotic safety constrains the Horndeski Lagrangian.
\par\medskip

We further identify a second ``twin'' Horndeski trajectory corresponding to a positive de Sitter (dS) value of the cosmological constant at around $g\lambda\approx+2\times10^{-8}$.
This trajectory corresponds to an angle $\alpha$ slightly above the separatrix, approximately the mirror image of the AdS Horndeski trajectory.
As discussed earlier, every dS trajectory runs into the lambda-1/2 singularity, which is why the dS Horndeski trajectory is not displayed in Fig.~\ref{fig:IRValues02pi}.
Generally, one would thus expect that no dS path leads to sensible physical $k\to0$ (IR) data.
However, close to the separatrix, dS and AdS trajectories behave mirror-symmetrically until the onset of the singularity in the dS trajectories: changing the sign of $\lambda$ leaves the other couplings invariant.
We thus expect that the converged AdS Horndeski trajectory has a twin dS trajectory of the same $c_S/c_T$ value. 
Indeed, we identify a trajectory with $c_S/c_T\approx-1/2$ right before the onset of the singularity. 
The question of whether a different treatment of the off-shell pole in the metric propagator can give rise to a Horndeski trajectory with positive $\Lambda$ in the IR is left to future work.
For the remainder of this paper, we focus on the Horndeski trajectory with negative $\Lambda$.

\subsubsection{
Mechanism for the emergence of Horndeski gravity from asymptotic safety
}\label{subsec:FlowRegimes}

Our next goal is to better understand the mechanism which allows the ratio of non-minimal couplings to reach the Horndeski value. This clearly hinges on a growth of the ratio at intermediate scales between the Planck scale and IR scales for trajectories close to the separatrix, see Fig.~\ref{fig:IRValues02pi}.
To understand the mechanism behind this growth and the overall behavior of $c_S/c_T$ , we discuss the different regimes of the RG flow for trajectories close to the separatrix, see Fig.~\ref{fig:HorndeskiRGFlow}.

According to the physical direction of the flow, we start the discussion in the UV and move towards smaller $t$, into the IR.
For a better overview, we formulate this discussion in the RG ``time'' $t=\ln(k/k_0)$. A different choice of reference scale $k_0$ would thus shift this discussion by a constant in $t$. In practice, this is related to a different choice of $\epsilon$ in Eq.~\eqref{eq:InitalPointFlow}.
At large t, in the deep UV, the flow starts close to the fixed point, and all  dimensionless couplings are constant (quantum scale symmetry).
Around the Planck scale, at $t\approx5$ in Fig.~\ref{fig:HorndeskiRGFlow}, the flow exits the UV fixed-point regime, and strong renormalization effects kick in.
In the linear regime around the fixed point, this is controlled by the complex critical exponents in Eq.~\eqref{eq:CritExponents} and results in the rotation visible in Fig.~\ref{fig:StreamPlotHorndeski}.
The Horndeski trajectory lies close to the separatrix, which means that it subsequently approaches the Gaussian fixed point.
This corresponds to the region $-1\lesssim t\lesssim 3$ in Fig.~\ref{fig:HorndeskiRGFlow}.
Accordingly, it exhibits Gaussian fixed-point scaling, where both $g$ and $\lambda$ scale as $k^{-2}\propto \exp(-2t)$.
Once the flow is close enough to the fixed point, at $t\approx2.5$ in Fig.~\ref{fig:HorndeskiRGFlow}, the dimensionless ratios $c_S/g$ and $c_T/g$ start scaling linearly in $t$, which corresponds to logarithmic growth in $k=k_0 e^t$. 

To understand the origin of the logarithmic scaling, without which $c_S/c_T$ could not reach the Horndeski ratio, we consider the structure of the beta functions.
From $\beta_g$, we identify that below the Planck scale, we can set $g=G_Nk^2$.
We plug this into the beta functions of the non-minimal couplings and expand around small $k$ to obtain
\begin{align}\label{eq:cSBetaExpanded}
    \beta_{c_S}&=2c_S-\frac{(2+3\lambda-4\lambda^2)G_Nk^2}{6\pi(1-2\lambda)^3}+O(k^4),
    \\ \label{eq:cTBetaExpanded}
    \beta_{c_T}&=2c_T-\frac{G_Nk^2}{6\pi(1-2\lambda)^2}+O(k^4).
\end{align}
The quantum corrections arises from the fact that the couplings in question obey shift symmetry and $\mathbb{Z}_2$ symmetry, which is the full symmetry of the scalar kinetic term. Accordingly, the non-minimal couplings are induced by metric fluctuations, even if they were set to zero. This is encoded in terms $\sim g$, which are independent of $c_S$ and $c_T$, respectively. 
\\
This mechanism, by which asymptotic safety for gravity prevents a fixed point that contains no matter interactions, is by now well-established \cite{Eichhorn:2011pc,Eichhorn:2012va, Meibohm:2016mkp,Eichhorn:2017eht,Christiansen:2017gtg,Eichhorn:2017sok,Eichhorn:2019yzm,deBrito:2021pyi,Laporte:2021kyp, deBrito:2020dta,deBrito:2023myf,deBrito:2023kow, Eichhorn:2024wba,deBrito:2025nog,Eichhorn:2026euv}. It is tied to the interesting question, whether asymptotic safety evades the no-global-symmetries conjecture, by only generating interactions that satisfy global symmetries of the matter kinetic terms, see \cite{Eichhorn:2020sbo,Eichhorn:2022gku,Eichhorn:2024rkc,Basile:2025zjc} for discussions.

On the separatrix, we can further use that $\lambda=\Lambda_0k^2$, as required to approach the Gaussian fixed point.
This leads to
\begin{align}
    \beta_{c_S}&=2c_S-\frac{G_Nk^2}{3\pi}+O(k^4),
    \\
    \beta_{c_T}&=2c_T-\frac{G_Nk^2}{6\pi}+O(k^4).
\end{align}
Using the expansion of the beta functions for the non-minimal couplings on the separatrix up to quadratic order in $k^2$, we find the approximate solutions
\begin{align}
    \frac{c_S}{g}&\approx c_{S,0}-\frac{1}{6}\log(G_Nk^2)\label{eq:cS_Separatrix}
    \\ \label{eq:cT_Separatrix}
    \frac{c_T}{g}&\approx c_{T,0}-\frac{1}{12}\log(G_Nk^2),
\end{align}
where $c_{S,0}$ and $c_{T,0}$ are integration constants that encode the UV behavior.
Higher-order contributions to the beta functions do not change this qualitative behavior.
This explains the logarithmic divergence of the non-minimal couplings close to the Gaussian fixed point observed in Fig.~\ref{fig:RGflowCouplingsHorndeski}.
Similar logarithmic scaling has been observed in other higher-order couplings \cite{Knorr:2024yiu, Eichhorn:2024wba,Knorr:2026vax}.
By dividing Eq.~\eqref{eq:cS_Separatrix} by Eq.~\eqref{eq:cT_Separatrix}, we find that exactly on the separatrix, $c_S/c_T\to2$ in the IR, which means that the spike in Fig.~\ref{fig:IRValues02pi} is bounded by 2 from above.

At the turning point, $t\approx-1$ in Fig.~\ref{fig:HorndeskiRGFlow}, the behavior of $\lambda$ changes from Gaussian scaling $\lambda\propto k^2$ to classical scaling $\lambda\propto k^{-2}$.
The turning point determines the scale at which the flow departs from the separatrix and starts moving outwards. 
By definition, the closer a trajectory is to the separatrix, the lower the RG ``time'' of the turning point.
The logarithmic scaling of the non-minimal couplings continues in the intermediate regime $-6\lesssim t\lesssim -1$, while the trajectory is still close enough to the Gaussian fixed point, such that $\lambda$ remains negligible.
When $\lambda$ becomes negative enough, the quantum contributions in Eqs.~\eqref{eq:cSBetaExpanded} and \eqref{eq:cTBetaExpanded} are suppressed, and the logarithmic scaling ends. 
Then, in the deep IR, the beta functions only contain the canonical terms and all dimensionless ratios freeze out, which means that the dimensionful couplings become constant.
\\
In summary, the scale dependence of the coupling ratio $c_S/c_T$ displayed in Fig.~\ref{fig:IRValues02pi} hinges on the logarithmic scaling close to the Gaussian fixed point. In the logarithmic scaling regime, $c_S/c_T$ increases from the fixed-point value of approximately $-3.5$. Rather non-trivially, the maximum value that can be achieved for the ratio is $c_S/c_T=2$, which is above the Horndeski ratio, while all values inbetween can be met.

This is by no means a guaranteed outcome. A priori, it would have been possible to find that the Horndeski ratio is simply not within the ``landscape'' associated to the asymptotically safe scaling regime. However, in our case, we find that the Horndeski ratio is ``in the landscape''. To reach this specific point in the landscape, a long logarithmic scaling regime is needed. This regime makes it possible to achieve the Horndeski ratio $c_S/c_T=-1/2$. In turn, a long logarithmic scaling regime is enabled by trajectories very close to the separatrix. As a consequence, we can ``sacrifice'' a relevant direction, associated to $\Lambda$, to select a trajectory that is exactly as close to the separatrix as is needed to achieve the Horndeski ratio in the IR.

\section{Effective asymptotic safety and Horndeski gravity}\label{sec:EffectiveASQG}
The aim of the asymptotic safety program is to establish a predictive quantum theory of gravity, which describes gravity at much smaller distance scales than is consistently possible in classical GR.
This goal can be fulfilled without the expectation that asymptotic safety is the fundamental theory, which holds up to arbitrarily small scales.
Instead, an asymptotically safe fixed point can render a QFT predictive, even if that QFT is still an EFT with a UV cutoff, $k_{\text{UV}}$, beyond which which an even more fundamental description sets in.
This could be, for example, string theory, loop quantum gravity or some other description that has yet to be developed; see \cite{Eichhorn:2026uqj} for a summary of concrete indications for points of contact between these different approaches and asymptotic safety.
In such an ``effective asymptotic safety'' scenario \cite{Percacci:2010af,deAlwis:2019aud,Held:2020kze}, quantum scale symmetry is approximately fulfilled between the Planck scale and some new UV scale $k_{\text{UV}}$.
In this section, we want to explore the implications of an effective asymptotic safety scenario on the previously found Horndeski trajectory.
\par\medskip

\begin{figure}
    \centering
    \includegraphics[width=0.8\linewidth,clip=true, trim=1.5cm 0cm 3cm 0cm]{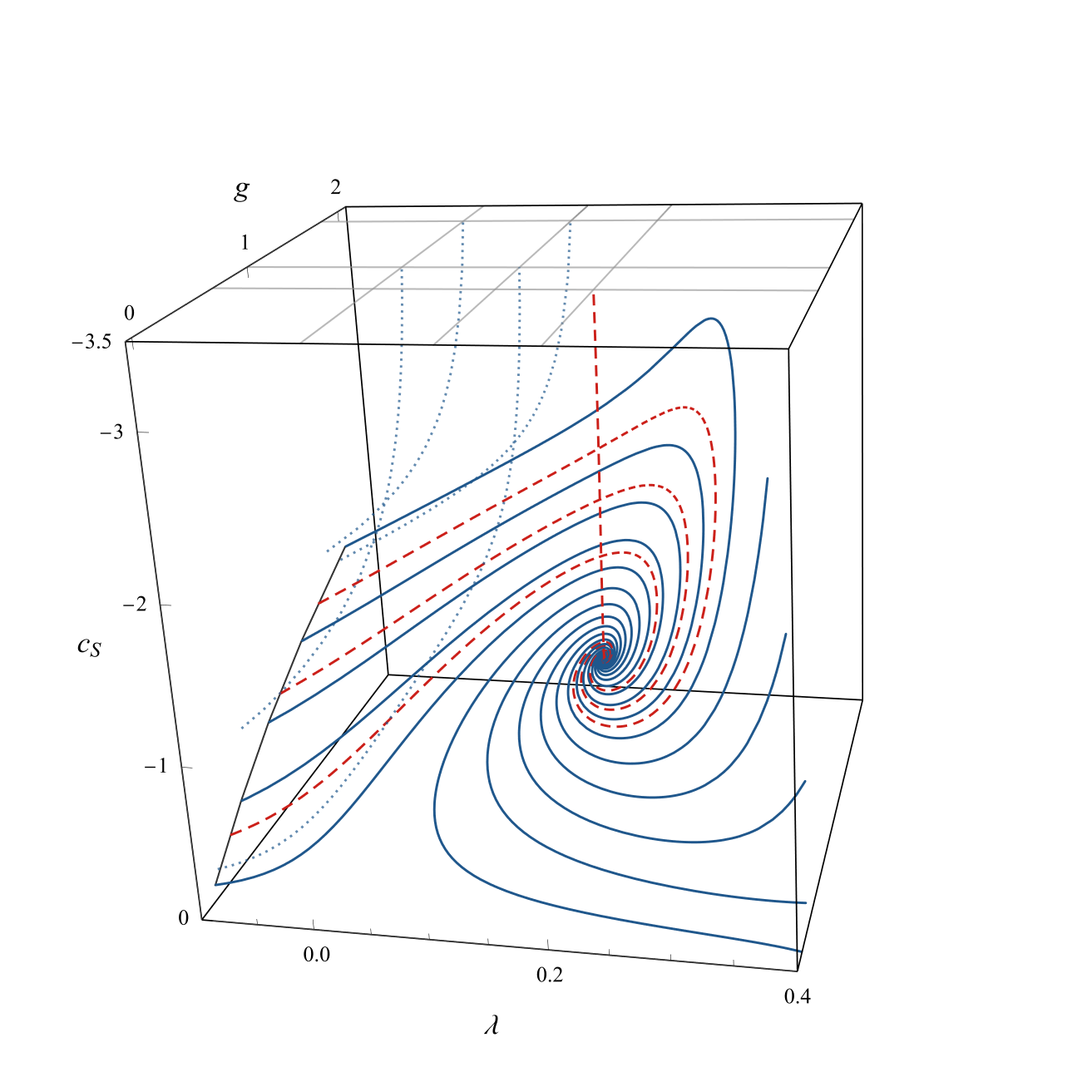}
    \caption{We show the RG flow projected onto the three-dimensional $(g,\lambda,c_S)$ cube around the fixed point.
    The blue trajectories, calculated with $\beta=0$, are true fixed-point trajectories, as in the two-dimensional version in Fig.~\ref{fig:glambdaStreamPlot_0-2pi}.
    They map out the UV critical surface of the fixed point.
    The three dashed red trajectories are 
    trajectories realizing effective asymptotic safety.
    They do not emanate from the fixed point in the UV, as can be seen from the dashed red line above the fixed point, but instead pass by close to it.
    These trajectories approximately lie in the critical surface, which explains why, in this system, the predictions obtained in the effective and the fundamental asymptotic safety scenario are near indistinguishable.
    The dotted blue lines are trajectories that do not exhibit fixed-point scaling. Nevertheless, they are attracted to the UV critical surface in the IR.
    }
    \label{fig:effectiveASQGStreamPlot}
\end{figure}

Let us first discuss how the effective asymptotic safety perspective changes the flow of the couplings.
The key difference is that the RG flow in effective asymptotic safety does not originate from the fixed point in the deep UV. 
This is displayed by the dashed red trajectories in Fig.~\ref{fig:effectiveASQGStreamPlot}.
The more fundamental theory predicts some initial point $P$ of the RG flow, from where the flow approaches the fixed point. It is therefore not reasonable to interpret the flow above the unknown scale $k_{\text{UV}}$; instead, one likely encounters a Landau pole or strong-coupling regime in some couplings which require or hint at the presence of new physics.
The flow towards the IR stays close to the Reuter fixed point in the intermediate regime between $k_{\text{UV}}$ and the Planck scale, where it exhibits approximate quantum scale symmetry, as in fundamental asymptotic safety.
\\
In terms of the eigensystem of the fixed point, the initial point $P$ thus contains not only a step into a relevant direction away from the fixed point, but also a step into an irrelevant direction.
Along the irrelevant directions, the RG flow is attracted towards the fixed point, while the relevant directions push for a departure from the fixed point regime.
At around the Planck scale, the flow departs from the vicinity of the fixed point and flows into the IR, as in the setting of ``fundamental asymptotic safety''.
\\
A powerful consequence of such a scenario is that the predictions generated by the asymptotically safe fixed point and its critical hypersurface carry over to the more fundamental theory, see, e.g., \cite{Basile:2021krr,Knorr:2024yiu}. If a set of distinct fundamental theories are all asymptotically safe, then effective asymptotic safety generates universal predictions for these.

To harness the predictive power of the fixed point for effective asymptotic safety in this way, the initial point of the flow has to be close enough to the fixed point to generate a regime of approximate quantum scale symmetry.
Generally, the closer the initial point is to the \emph{IR} critical surface of the fixed point (i.e., to the surface spanned by the \emph{irrelevant} directions), the closer the physical predictions of effective asymptotic safety are to the predictions of fundamental asymptotic safety.
Usually, due to the funneling nature of the fixed point, when the RG flow covers at least a few orders of magnitude in a nearly quantum scale symmetric regime, the different IR predictions are near indistinguishable \cite{deAlwis:2019aud}.
This underlines the universality of the predictive mechanism of asymptotic safety.
\par\medskip

For our analysis of all effectively asymptotically safe trajectories in our gravity-scalar system, we treat the entire four-dimensional coupling space close to the fixed point as possible initial conditions of the flow.
This includes the two relevant and the two irrelevant directions.
We therefore generalize the initial point of the flow from Eq.~\eqref{eq:InitalPointFlow} to 
\begin{align}\label{eq:InitialPointFlowEffectiveASQG}
\begin{split}
    P=\text{FP}+\epsilon\Big[
    \cos(\beta/2)&\left(\cos(\alpha)\widetilde V_1+\sin(\alpha)\widetilde V_2\right)
    +\\
    \sin(\beta/2)&\left(\cos(\delta)\widetilde V_3+\sin(\delta)\widetilde V_4\right)\Big].
\end{split}
\end{align}
As in Eq.~\eqref{eq:InitalPointFlow}, $\widetilde V_{1,2}$ are the real projections of the pair of relevant eigenvectors in Eq.~\eqref{eq:Eigenvectors} and $\widetilde V_{3,4}$ the real projections of the pair of irrelevant eigenvectors in Eq.~\eqref{eq:Eigenvectors34}.
The parametrization of the angles in Eq.~\eqref{eq:InitialPointFlowEffectiveASQG} retains the interpretation of $\alpha$ as the angle between the relevant directions.
Given a combination of the angles $(\beta,\delta)$, the distribution in terms of $\alpha$ can be compared to the IR results of the previous section in Fig.~\ref{fig:IRValues02pi}.
$\beta$ continuously interpolates between relevant and irrelevant directions.
For $\beta=0$, we turn off the irrelevant directions and retain only the relevant directions, recovering our previous results.
For $\beta=\pi$, the initial step is instead along a purely irrelevant direction.
Such a flow converges exactly onto the fixed-point value.\footnote{
The critical hypersurface is generally curved. For finite step size $\epsilon$, a step into an irrelevant direction therefore also contains a small relevant part.
What happens in practice after an initial step into an irrelevant direction is the following: flowing towards the fixed point, the irrelevant parts of the initial step shrink until the tiny, relevant parts become significant.
}
The step size $\epsilon$ directly influences ``how effective'' the flow is, so over how many orders of magnitude it (nearly) realizes quantum scale symmetry.
To a lesser degree, this also depends non-trivially on the angles.

Given a value of $\epsilon$, we can now scan grids of angles $(\alpha,\beta,\delta)$.
Choosing $\epsilon=10^{-4}$ and $\epsilon=10^{-3}$, we find multiple orders of magnitude of quantum scale symmetry in the UV for most angles.
For $\epsilon=10^{-2}$, $\epsilon=10^{-1}$, or even larger values, we exit the regime of effective asymptotic safety.
To connect to previous results, for a set of $(\beta,\delta;\epsilon)$, we repeat the procedure of Fig.~\ref{fig:IRValues02pi}: we integrate into the IR and evaluate the IR statistics as a function of $\alpha$.
\\
All analysis of the four $\epsilon$ sizes $\{10^{-4},10^{-3},10^{-2},10^{-1}\}$ find that for every tested combination of $(\beta,\delta)$ the physical (IR) predictions do not deviate significantly from the previous fundamental asymptotic safety analysis.\footnote{This analysis includes, among others, a $16\times 12\times12$ grid search of the angles for $\epsilon=10^{-1}$, where no significant deviations from the fundamental asymptotic safety results were found, even though $\epsilon=10^{-1}$ is already large enough to include almost no UV fixed point scaling.
}
In particular, we find the same  log-scaling near the separatrix and no additional trajectories that lead to physical predictions of $c_S/c_T>-1$.

The funneling of the fixed point is too strong to allow for a significant change in physical predictions.
We conclude that the prediction of the relation between the asymptotically safe scalar-tensor theory and Horndeski gravity discussed in Sec.~\ref{sec:Results1} remains unchanged within the effective asymptotic safety scenario. 
This implies that any quantum-gravity theory that serves as a UV completion of the asymptotic-safety scenario (and therefore provides UV initial conditions at $k_{\rm UV}$ close enough to the IR critical surface of the fixed point), achieves the Horndeski ratio in the IR, provided the relevant direction associated to $\Lambda$ is tuned to the same value of $\Lambda$ in the IR as in the fundamental asymptotic safety scenario.
\par\medskip

It is an expected outcome that trajectories that are close enough to the IR critical surface of a fixed point result in similar predictions for the IR. It is less expected that such universal predictions are also realized for starting points which do not appear to be close to the IR critical surface and which do not realize an intermediate, nearly scale invariant regime.\footnote{It may be possible to define a metric in the space of couplings that measures distances in an appropriate way, such that UV initial conditions which result in similar IR predictions are close in that metric. Here, we base our estimation of whether UV initial conditions are ``close'' or ``far'' by the criterion of whether the resulting trajectories exhibit a regime during which they realize approximate scale symmetry.}
We trace this back to the size of the critical exponents.
Consider again the flow resulting from the linearized beta functions in Eq.~\eqref{eq:CouplingExpandedFP}, where  the real parts of the critical exponents are $\text{Re}(\theta_{\text{rel}})\approx1.5$ for the relevant directions and $\text{Re}(\theta_{\text{irrel}})\approx-2.6$ for the irrelevant directions.
Formulated in the RG time $t$, the scaling of the flow in the direction $V^J$ with associated critical exponent $\theta_J$ reads $\exp(-\theta_J\, t)$.
Because $|\text{Re}(\theta_{\text{rel}})|<|\text{Re}(\theta_{\text{irrel}})|$, the exponential scaling entails that, close to the fixed point, the irrelevant directions flow onto the fixed point much faster than the relevant directions can flow away from it.
This explains why an additional step into an irrelevant direction cannot lead to a significant change in physical IR predictions: it is undone before the flow exits the fixed-point regime, where the step could have non-trivial impact.
By the same argument, starting outside the critical hypersurface, the RG flow orthogonal to the critical hypersurface is much faster than the flow parallel to it. 
In other words, the critical hypersurface is highly IR attractive.
This explains why even RG trajectories that do not feature an intermediate, nearly scale invariant regime can still result in very similar IR predictions, as displayed by the dotted blue trajectories in Figure~\ref{fig:effectiveASQGStreamPlot}.

We conclude with a generalization: in effective asymptotic safety, for $|\text{Re}(\theta_{\text{rel}})|\ll|\text{Re}(\theta_{\text{irrel}})|$, the ``funneling effect'' of the fixed point is so strong that the possible IR values do not significantly deviate from those of fundamental asymptotic safety. Accordingly, the universal predictions associated to the UV critical surface of the fixed point may be important much more generally than has previously been appreciated.

\section{Radiative stability of Horndeski coupling ratios under gravitational fluctuations}\label{sec:RadiativeStability}

In this section, we discuss the radiative stability of the classical Horndeski theory in Eq.~\eqref{eq:ClassicalHorndeskiAction}, independently of any UV completion and, in particular, independent of asymptotic safety.
Radiative stability describes the property of a classical theory to remain consistent when accounting for quantum corrections. In particular, radiative stability is an important requirement in settings where a classical theory is defined by special ratios of couplings which are not protected by symmetries, as is the case for Horndeski gravity.
\par\medskip
 
Horndeski gravity is often investigated at a purely classical level, without accounting for quantum fluctuations.
However, quantum fluctuations can generate additional interactions that are not present in the classical Lagrangian: not all choices of classical actions are consistent at the quantum level.
In the case of the theory in Eq.~\eqref{eq:ClassicalHorndeskiAction} with Horndeski term $ G^{\mu\nu}\partial_\mu\phi\partial_\nu\phi$, we a priori expect that quantum fluctuations generate the operators $C_T R^{\mu\nu}\partial_\mu\phi\partial_\nu\phi$ and $C_S R(\partial\phi)^2$ separately.
However, only for the specific coupling relation $C_S/C_T=-1/2$ does the theory remain a member of the Horndeski family.
A nontrivial test of radiative stability in this Horndeski theory, therefore, consists of analyzing the change of the ratio $C_S/C_T$ under the RG flow in the larger theory space that contains $C_S$ and $C_T$ separately. We will do so without requiring a UV completion; accordingly, RG trajectories are not complete. To show a lack of radiative stability, it is already sufficient to check whether the RG flow deviates from $C_S/C_T=-1/2$ after just a short range of scales, independently of whether or not the RG trajectory in question is complete.

Within an EFT context, radiative stability does not strictly require that
all Horndeski-breaking terms are completely absent.
It requires that if any such terms appear, they are strongly suppressed, such that they can effectively be ignored at EFT scales.
In many studies of radiative stability, this can be controlled through symmetry properties, which might prohibit or suppress the generation of certain symmetry-breaking terms. 
In the case of Horndeski gravity, there is no symmetry that enforces $C_S/C_T=-1/2$, as this property ``only'' ensures the absence of additional degrees of freedom.\footnote{In an order-reduced setting, in which new degrees of freedom are not encoded in higher-order equations of motion, but are made explicit through additional fields, the symmetry group is not changed by a change in the number of degrees of freedom, but new representations of the symmetry group are added through the new fields. It is an interesting question whether this may result in non-renormalization effects.} 
Therefore, a priori, there is no reason to expect quantum corrections to respect the Horndeski ratio.
Against this background, understanding whether or not Horndeski gravity is radiatively stable, is a pressing issue.
\\
The radiative stability of various theories within the Horndeski framework has previously been investigated, e.g., in \cite{Brouzakis:2013lla,Pirtskhalava:2015nla,Arbuzov:2017nhg,Latosh:2018xai,Noller:2018eht,Santoni:2018rrx,Latosh:2020jyq,Heisenberg:2020cyi,Latosh:2021usy}, sometimes using symmetries (e.g., shift-symmetry) in subsectors of the theory to achieve the desired suppression of newly generated interactions.
However, to the best of our knowledge, the two non-minimal couplings $C_S$ and $C_T$ of this work have not yet been studied with respect to the radiative stability of the Horndeski ratio.
\par\medskip

\begin{figure}[t]
    \centering
    \includegraphics[width=0.8\linewidth]{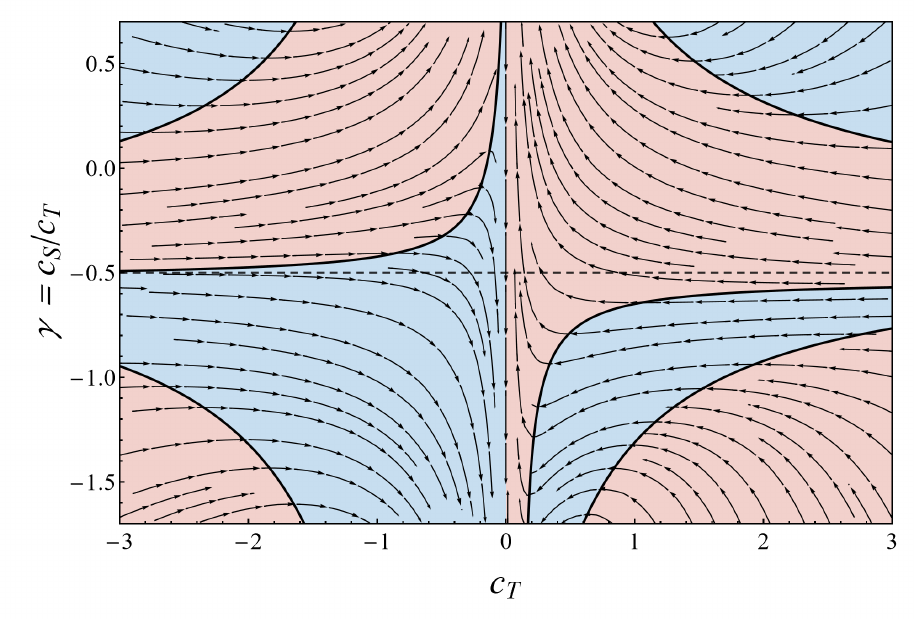}
    \caption{\label{fig:radiativestability} We show the RG flow in the $(c_T,\gamma)$-plane at constant $(g,\lambda)$.
    In the red areas, the flow towards the IR leads to an increase in $\gamma$ ($\beta_\gamma>0$) and in the blue areas to a decrease in $\gamma$ ($\beta_\gamma<0$). On the black lines the value of $\gamma$ is locally preserved ($\beta_\gamma=0$).
    The Horndeski value $\gamma=-1/2$ is highlighted by the dashed line.
    \\
    We can identify that, in general, the flow lines cross the $\gamma=-1/2$ line, driving the theory away from the Horndeski value.
    The Horndeski theory in Eq.~\eqref{eq:ClassicalHorndeskiAction} is not in general radiatively stable.
    However, outside of $|c_T|\lesssim 0.6$, the flow lines are nearly parallel to the Horndeski line, indicating that the violation of the Horndeski condition may remain small.
    }
\end{figure}

We consider the RG flow of the dimensionless couplings $c_S$ and $c_T$ to study whether the Horndeski ratio $C_S/C_T=c_S/c_T=-1/2$ is radiatively stable.
To that end, it is useful to describe the system in a different  basis in the space of couplings.
We introduce the coupling ratio
\begin{equation}
\gamma= \frac{c_S}{c_T},
\end{equation}
and derive its beta function,
\begin{equation}
\beta_{\gamma} = \frac{\beta_{c_S}}{c_T} - \frac{c_S}{c_T^2}\beta_{c_T}.
\end{equation}
Strict radiative stability of Horndeski gravity would require that
\begin{equation}\label{eq:radiativestability}
\beta_{\gamma}\Big|_{\gamma=-\frac{1}{2}}=0,
\end{equation}
i.e., quantum fluctuations do not generate any deviation from the Horndeski ratio.
As we can see in Fig.~\ref{fig:radiativestability}, this is, in general, not realized.
Instead, the flow departs from the Horndeski ratio, leading to higher-order equations of motion.

We do, however, find the following unexpected result, which hints at further-reaching implications, if similar results  can also be found for larger truncations:
at large $|c_T|$,  the flow becomes approximately parallel to the Horndeski line, cf.~Fig.~\ref{fig:radiativestability}, indicating approximate radiative stability.
\\
We study the asymptotic behavior of $\beta_\gamma$.
An expansion of the beta function around large $c_T$ for fixed $\gamma$ leads to
\begin{equation}\label{eq:betagamma_largecT}
\beta_{\gamma} \xrightarrow{c_T \rightarrow \pm \infty}
 \left(g^2 \frac{\left(1+ 6 \gamma + 8 \gamma^2 \right)^2}{168 \pi \left(g-12 \pi (1-2 \lambda)^2+ 10 g\,\lambda \right)} \right) c_T^3.
\end{equation}
In this limit, the beta function has a zero not only at $\gamma=-1/4$ but also for the Horndeski value $\gamma=-1/2$.
Strictly at $\gamma=-1/2$, the $\mathcal{O}(c_T^2)$ terms become important, which entails that the full beta function is not exactly zero at this value of $\gamma$.
This explains why the RG flow in Fig.~\ref{fig:radiativestability} becomes near-horizontal, but not perfectly horizontal, for large $|c_T|$ close to $\gamma=-1/2$.
\\
We consider it a rather surprising and nontrivial result that there exists a regime in which the Horndeski ratio constitutes a special value for the RG flow and that locally, radiative stability of the Horndeski ratio is approximately realized. 
We stress that this behavior is even independent of the values of $\lambda$ and $g$.
One may therefore interpret this result as a tentative indication that, within the larger space of scalar-tensor theories, the absence of additional degrees of freedom entailed by the second-order nature of Horndeski gravity, may be (at least partially) protected under quantum effects.
We further consider our results a strong motivation for a more comprehensive study accounting for further interactions, which is beyond the scope of the present paper.

\section{Conclusion and outlook}
An asymptotically safe scalar-tensor theory is not a priori expected to give rise to a Horndeski theory as its low-energy EFT. There is no obvious reason for why additional couplings, leading to higher-order equations of motion and not constrained by additional symmetries, should vanish. 
However, higher order equations of motion are related to additional propagating gravitational degrees of freedom which may be difficult to reconcile with observations and theoretical consistency.
At the same time, there is already evidence that asymptotic safety manages to avoid extra degrees of freedom in the purely gravitational sector because the graviton propagator shows no indication of extra poles \cite{Fehre:2021eob,Pawlowski:2025etp,Knorr:2026jcg,Assant:2026dca}.
While these results have been achieved in truncations of the full dynamics, and therefore are by no means a proof of the absence of extra degrees of freedom, they constitute highly non-trivial results.
The full quantum-gravitational graviton propagator is a non-trivial function of the momentum and not necessarily well-approximated by the classical GR transverse traceless propagator.
For scalar matter, the situation is less clear-cut, based on an analogous study of the scalar propagator \cite{Kher:2025rve}.

Here, we study an alternative mechanism in which special values of couplings reduce the order of the equations of motion already in a setting with finitely many higher-order derivatives.
Conceptually, there is a relation of our work to the study in \cite{Becker:2017tcx}, in which additional modes are absent because their mass achieves a special, namely arbitrarily large, value in the IR.
We consider two couplings, $C_S$ and $C_T$, generated by asymptotic safety that, in general, render the equations of motion higher order.
The theory only reduces to second order equations of motion for one particular ratio of the couplings, where it is a Horndeski theory.
We find that for a particular, small value of the cosmological constant $\Lambda$, the Horndeski ratio can be realized in asymptotic safety. At the same time, the values of $C_S$ and $C_T$ are fixed individually at this point.
We do not reach a general Horndeski Lagrangian, but one with a specific coupling value. 
\\
Formulated differently, we find that, for a massless, $\mathbb Z_2$-symmetric scalar-tensor theory, the Horndeski theory in Eq.~\eqref{eq:ClassicalHorndeskiAction} lies in the ``landscape'' of asymptotic safety at one specific value of the coupling. 
In the truncation employed in this work, the coupling value comes out as $C=-0.82\,[m_P^{-2}]$, after a sign change due to the Wick rotation to Lorentzian signature.
This is a nontrivial result related to the logarithmic scaling of massless modes, as explained in Sec.~\ref{subsec:FlowRegimes} and displayed in Fig.~\ref{fig:IRValues02pi}. At the same time, $C\ne-0.82\,[m_P^{-2}]$ lies in the ``swampland'' of asymptotically safe gravity. 
We stress again that the above statements, of course, all hold within the truncation defined in  Eq.~\eqref{eq:ClassicalHorndeskiAction}. This result also provides evidence for the expectation that the ``landscape'' of asymptotically safe theories is relatively small, which is related to the strong predictive power of asymptotic safety.

In terms of the mechanism underlying our result, it is the ``sacrifice'' of the cosmological constant as a free parameter that allows us to achieve the coupling ratio corresponding to Horndeski gravity in the IR.
Such a mechanism can clearly not be generalized arbitrarily because asymptotic safety contains only a small number of free parameters.
The question of whether Horndeski gravity can be ruled out as the EFT approximation to asymptotically safe scalar-tensor theories becomes even more nontrivial in larger truncations, and this work highlights the importance of further investigation of this issue.
The next nontrivial test could involve adding further terms which also reduce to second-order equations of motion only in specific coupling relations.
This allows to see if our result can persist when no further relevant directions can be ``sacrificed'' to obtain a desired value for an irrelevant direction.
Our results also require corroboration in Lorentzian signature in the future.
In a more general setting, one could further generalize the question to massive scalars and study DHOST theories instead of Horndeski gravity.

The present, as well as analogous future studies, are of interest to several research communities.
For cosmology, where model-building for dynamical dark energy often proceeds within the framework of Horndeski gravity, such studies can either support or discourage the use of this framework; similar statements apply to other modified-gravity settings. For asymptotic safety, such studies can further elucidate the number and nature of propagating degrees of freedom. This has consequences both for the fundamental consistency as well as the potential phenomenological consequences of the theory. For the swampland program, such studies single out concrete theories that lie in the swampland or the landscape of asymptotic safety and therefore provide concrete examples with which to sharpen the distinction of various swamplands, and e.g., gain insight into whether the relative swamplands \cite{Eichhorn:2024rkc} of asymptotic safety and, e.g., string theory, have overlap.

We have further established that in an effective asymptotic safety scenario, the predictions for $C_T$ and $C_S$, and thereby the Horndeski ratio, remain unchanged.
The mechanism behind this is that  the irrelevant directions are more attractive than the relevant directions are repulsive in the vicinity of the fixed point.
We found that this makes the critical surface strongly IR attractive, even away from the fixed point. This has consequences for Horndeski gravity beyond asymptotic safety, because it implies that even with other UV completions, only the value $C=-0.82\,[m_P^{-2}]$ can be achieved in Horndeski gravity, if the RG flow does not start too far away from the IR critical hypersurface of the fixed point in the UV.

Lastly, we have investigated the radiative stability of the Horndeski ratio, independently of asymptotic safety as a UV completion or extension of the system.
As expected, there is no strict radiative stability and, generally, quantum gravitational corrections lead to deviations from the Horndeski property.
However, for certain parameter ranges we have found approximate radiative stability of the Horndeski ratio, as long as the absolute value of the scalar couplings do not become too small.
This unexpected result also motivates a more general analysis in the future. If Horndeski gravity can be shown to be radiatively stable for certain regimes of coupling values, this provides a theoretically preferred range of couplings, which can in turn inform phenomenological studies.

Overall, we conclude that radiative stability, or even asymptotic safety of Horndeski gravity \emph{has not been ruled out} with our study, as would have been our expectation. Future investigation of these questions therefore is warranted.

\acknowledgments
We thank B.~Knorr for discussions.
We acknowledge the European Research Council's (ERC) support under the European Union’s Horizon 2020 research and innovation program Grant agreement No.~101170215 (ProbeQG).
This work is partially funded by the Deutsche Forschungsgemeinschaft (DFG, German Research Foundation) under Germany’s Excellence Strategy EXC 2181/1 - 390900948 (the Heidelberg STRUCTURES Excellence Cluster).
F.~W.~is supported by a scholarship of the German Academic Scholarship Foundation (Studienstiftung des deutschen Volkes).

\appendix

\section{Test of robustness: extended truncation
}\label{app:ExtendedTruncation}

To check the robustness of our results, we cross-check with a more extended truncation, which includes the flow of the scalar anomalous dimension as well as the square of the scalar kinetic term:

\begin{align}\label{eq:ExtendedTruncation}
\begin{split}
    \Gamma_{\text{ext}}=\int d^4x\sqrt{\det g_{\mu\nu}}\Big(
    \frac{2\Lambda-R}{16\pi G}
    +\frac{Z}{2}(\partial\phi)^2
    +Z^2C_K(\partial\phi)^4
    \\
    +C_{T} R^{\mu\nu}\partial_\mu\phi\partial_\nu\phi
    +C_{S} R(\partial\phi)^2
    \Big).
\end{split}
\end{align}

In this truncation, $Z$ and $C_K$ are also scale dependent couplings.
The additional kinetic-square operator is mass dimension eight.
It is thus the next scalar operator generated by the RG flow by count of operator dimension.
Stability of the results under extension of the truncation is an important criterion of the convergence of the ansatz.
\\
Again, we can work with the beta functions provided by \cite{Laporte:2021kyp} and confirm the existence of two zeros of the beta functions, one of which is a potential fixed point.\footnote{The potential fixed point carries an anomalous dimension around $-2.4$ and is thus on the verge of being a suitable fixed-point candidate.
The other zero of the beta functions has an anomalous dimension above 4 and flips the relevancy of a scalar coupling. This indicates a breakdown of the truncation and means the zero is not a suitable fixed point.}
We repeat the analysis with the additional irrelevant direction $C_K$ and find the same qualitative result:
For the two scalar operators to combine to the Einstein tensor such that the effective low energy action is a Horndeski theory, we require a small absolute value of the cosmological constant in the IR.

\bibliographystyle{JHEP}
\bibliography{references}

\end{document}